\documentclass[journal]{IEEEtran}
\usepackage{graphicx} % for pdf, bitmapped graphics files
\usepackage{graphics} % for pdf, bitmapped graphics files
\usepackage{subfigure}
\usepackage{algorithm}
\usepackage{algorithmic}
\usepackage{multirow}
\usepackage{amssymb}
\usepackage{amsmath}
\usepackage{xcolor}
\usepackage{mathrsfs}
\usepackage{booktabs}
\usepackage{cite}
\usepackage{caption}
\usepackage{stfloats}
\usepackage{amsthm}
\usepackage{nomencl}
\usepackage{verbatim}
\usepackage{color}

\usepackage{threeparttable}
\usepackage{amsmath}
\newtheorem{theorem}{\bf Theorem}

\usepackage{xstring}
\usepackage{xstring}
\usepackage{xpatch}
\makenomenclature

\patchcmd{\thenomenclature}
{\leftmargin\labelwidth}
{\leftmargin\labelwidth\itemindent 1em }
{}{}

\newcommand{\nomenclheader}[1]{%
	\item[\hspace*{-\itemindent}\normalfont\bfseries#1]}
\renewcommand\nomgroup[1]{%
	\IfStrEqCase{#1}{%
		{V}{\nomenclheader{Variables}}%   V - Variables
		{P}{\nomenclheader{Parameters}}% P - Parameters
		{S}{\nomenclheader{Sets}}% A - Acronyms
		{M}{\nomenclheader{Matrixes}}% A - Acronyms
		{I}{\nomenclheader{Indices}}
	}%
}
\begin{document}
\title{Joint Optimization for Coordinated Charging Control of Commercial Electric Vehicles Under Distributed Hydrogen Energy Supply}

\author{Teng~Long,~\IEEEmembership{Student Member,~IEEE,}
	and~Qing-Shan~Jia$^{*}$,~\IEEEmembership{Senior Member,~IEEE}% <-this % stops a space
\thanks{*This work was supported in part by the National Natural Science Foundation of China (No. 62073182), the National Key Research and Development Program of China (2016YFB0901900), the National Natural Science Foundation of China under grants (No. 61673229 and U1301254), and the 111 International Collaboration Project of China (No. B06002).}% <-this % stops a space
\thanks{T. Long, and Q.-S. Jia are with the Center for Intelligent and Networked Systems, Department of Automation, Tsinghua University, Beijing 100084, China (e-mail: lt17@mails.tsinghua.edu.cn; jiaqs@tsinghua.edu.cn).}%
\thanks{$^{*}$Q.-S. Jia is the corresponding author.}%
}


\markboth{Transactions on Control Systems Technology}%
{Shell \MakeLowercase{\textit{et al.}}: Bare Demo of IEEEtran.cls for IEEE Journals}

\maketitle

\begin{abstract}
The transition to the zero-carbon power system is underway accelerating recently. Hydrogen energy and electric vehicles (EVs) are promising solutions on the supply and demand sides. This paper presents a novel architecture that includes hydrogen production stations (HPSs), fast charging stations (FCSs), and commercial EVs. The proposed architecture jointly optimizes the distributed hydrogen energy dispatch and the EV charging location selection, and is formulated by a time-varying bi-level bipartite graph (T-BBG) model for real-time operation. We develop a bi-level iteration optimization method combining linear programming (LP) and Kuhn-Munkres (KM) algorithm to solve the joint problem whose optimality is proved theoretically. The effectiveness of the proposed architecture on reducing the operating cost is verified via case studies in Shanghai. The proposed method outperforms other strategies and improves the performance by at least 13\% which shows the potential economic benefits of the joint architecture. The convergence and impact of the pile number, battery capacity, EV speed and penalty factor are assessed.
\end{abstract}

\begin{IEEEkeywords}
Electric vehicle, hydrogen energy, bipartite graph, stochastic programming
\end{IEEEkeywords}
\printnomenclature

\IEEEpeerreviewmaketitle

\section{Introduction}
\IEEEPARstart{A}{s} an important development trend of the smart grid, zero-carbon power systems have drawn much attention around the world recently \cite{dominkovic2016zero}. Hydrogen energy and electric vehicles (EVs) are regarded as promising solutions to achieve this goal on the supply and demand side, respectively. The emissions of EVs strongly depend on their electricity generation mix for recharging and can be further reduced through renewable energy supply such as wind, photovoltaic and hydrogen energy. With the rapid development of EVs, the large-scale uncontrolled EV charging loads can add great stress to the distribution power network and cause congestion, power losses, and voltage deviations. Since EVs have significant elasticity in terms of charging, a reasonable scheduling control can save the overall operating cost, increase the renewable energy penetration and provide several ancillary services \cite{jia2020review}.

The existing methods on control of private EVs often face the privacy and security issues. However, the electrification and charging scheduling of commercial vehicles for passenger transportation (e.g., ride-hailing) are clear initial markets for EV fleet operation and operating cost reduction. The cost reduction can be done by two ways, one is to improve the electrification rate of vehicles, the other is to take fully usage of cheaper renewable energy. Many cities in China, America, and Europe have gradually achieved the electrification of taxis or other commercial vehicles \cite{wang2019tcharge}. Therefore, it is of great practical interest for the transportation network companies to schedule a fleet of commercial EVs for passenger transportation under hydrogen energy supply in their businesses. 

This paper studies the operation problem of transportation network companies, solving two major sub-problems jointly including the hydrogen energy dispatch and EV charging location selection. This problem is challenging due to the following difficulties. \emph{First}, the size of the solution space increases exponentially fast with respect to the number of EVs, which makes the solving process time-consuming. Therefore, a computationally feasible algorithm is in demand for real-time operation. \emph{Second}, many factors need to be integrated considered with the decision-making process, including the operating cost, road network topology, driving trajectories of EVs, and renewable energy output. \emph{Third}, the control decision is coupled in time. And the future information is uncertain.

Research on the control of charging stations and EV fleets have been active for years. Many works from charging station perspective focus on the planning stage, including the siting of charging stations \cite{zhang2017second} and the EV fleet sizing problem \cite{vazifeh2018addressing} to study their economic advantages. On the other hand, the charging control both for a single EV and for a fleet were studied recently to achieve different goals, such as battery healthy \cite{fang2016health}, peak procurement minimization\cite{moghaddass2019smart}, and valley filling \cite{liu2017decentralized}, just to name a few. DeForest et al. \cite{pflaum2017probabilistic} solved the charging stations management problem for the day-ahead market based on load forecasts and randomized algorithms. Morstyn et al. solved the problem with consideration of battery voltage rise and maximum power limitation, which are commonly neglected \cite{morstyn2020conic}. Driven by the need of state space reduction, event-based optimization \cite{long2017multi} and data-driven method \cite{li2017data} have been developed for a large-scale EV fleet charging operation.

Compared with day-ahead market, the real-time scheduling of EV fleet is more realistic and challenging. Assuming the private EVs are the price-takers, liu et al. \cite{liu2017dynamic} and Ghosh et al. \cite{ghosh2017control} developed the price mechanism and admission control to motivate EVs for off peak charging. Another way to solve this problem is to discretize the time into periods and transform the online problem into several offline optimization problems \cite{koufakis2019offline}. Heuristic and rule-based methods are proposed due to the high requirement of solving speed in real-time operation which may lack mathematical performance guarantee\cite{zheng2018novel, tucker2019online}. However, these works assume the arrival process and charging location of EVs are uncontrollable, while the controllable part is the charging power and time. Zhang et al. \cite{zhang2020power} studies the PEV routing problem using a second order cone programming model. Different from our paper, this work schedules the private EVs from the perspective of the social coordinator and did not consider the scheduling of renewable energy.

Compared with existing results, this paper studies the joint optimization problem for transportation network companies and advances the relevant literature by the following main contributions:

\emph{First}, we propose a novel architecture where a company owns the hydrogen production stations (HPSs), fast charging stations (FCSs) and commercial EVs for passenger transportation. The proposed architecture jointly optimizes the hydrogen energy dispatch and EV charging location selection at the same time. Compared with the architecture that considers only one of these issues, the HPS-FCS-EV architecture can obtain better performance on reducing the operating cost.

\emph{Second}, we propose a time-varying bi-level bipartite graph (T-BBG) model to formulate the architecture for the real-time urban charging scenarios. Based on the receding-horizon control framework, a bi-level iteration optimization algorithm is developed to solve the problem. The linear programming (LP) and extended Kuhn-Munkres (KM) algorithm are used for the hydrogen energy dispatch and EV charging location selection, respectively. The optimality of the proposed method is proved theoretically.

\emph{Third}, case studies based on real data in Shanghai are conducted to demonstrate the effectiveness of the proposed method. Compared with other strategies, the total operating cost of the proposed method is reduced by at least 13\% which shows the potential economic benefits of the joint architecture. The convergence and influences of various factors are analyzed. 

The remainder of this paper is organized as follows. Section \ref{section_for} gives the description and mathematical models of the HPS-FCS-EV architecture. We develop the T-BBG model in Section \ref{section_tbbg} and introduce the proposed bi-level iteration algorithm in detail in Section \ref{section_solution}. Numerical experiments are presented in Section \ref{section_case}. Section \ref{section_conclustion} concludes the paper.

\section{Problem Formulation}\label{section_for}
\begin{figure}[htb]
	\centering
	\includegraphics[width = 0.45\textwidth]{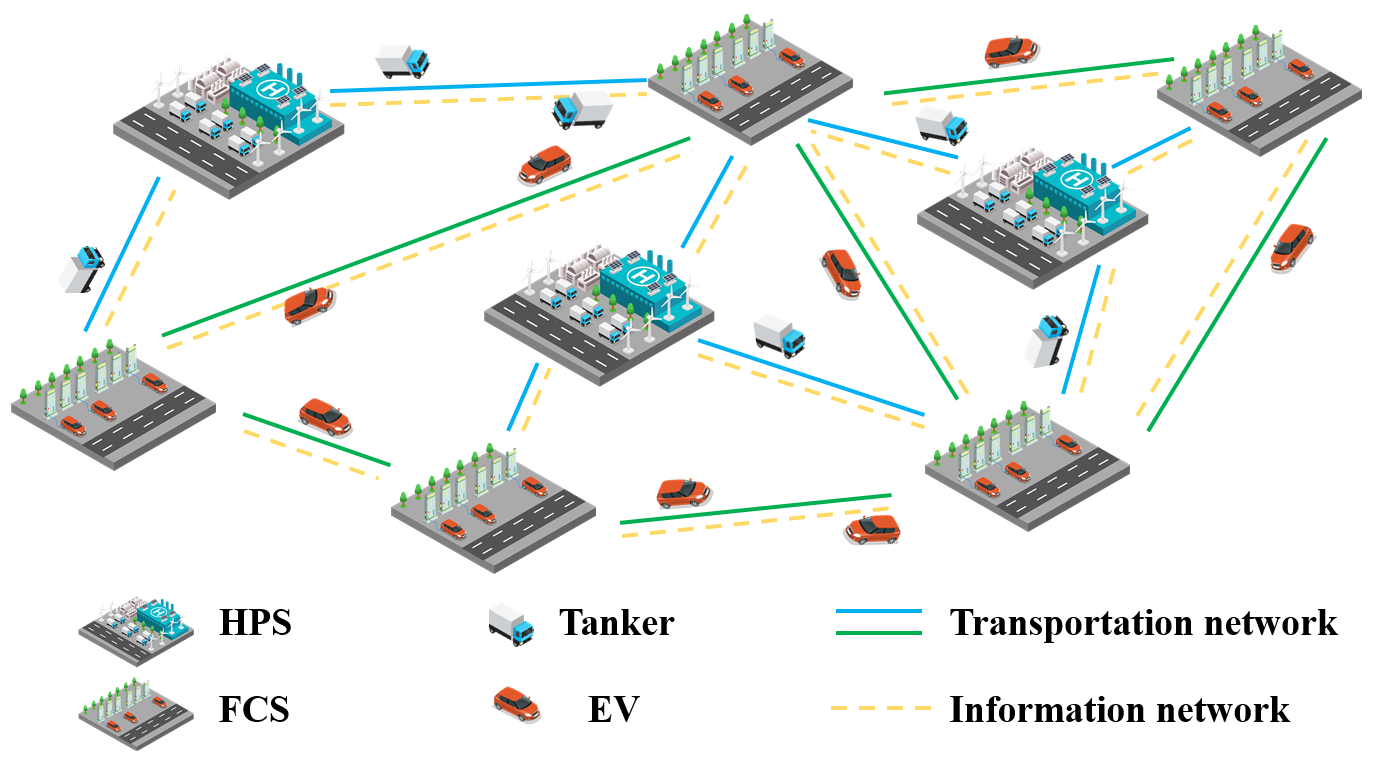}
	%	\captionsetup{margin=20pt,format=hang,justification=justified}
	\caption{The CPES system of a smart grid.}
	\label{fig:1}
\end{figure}

\begin{figure}[htb]
	\vspace{-0.2cm}
	\centering
	\includegraphics[width = 0.45\textwidth]{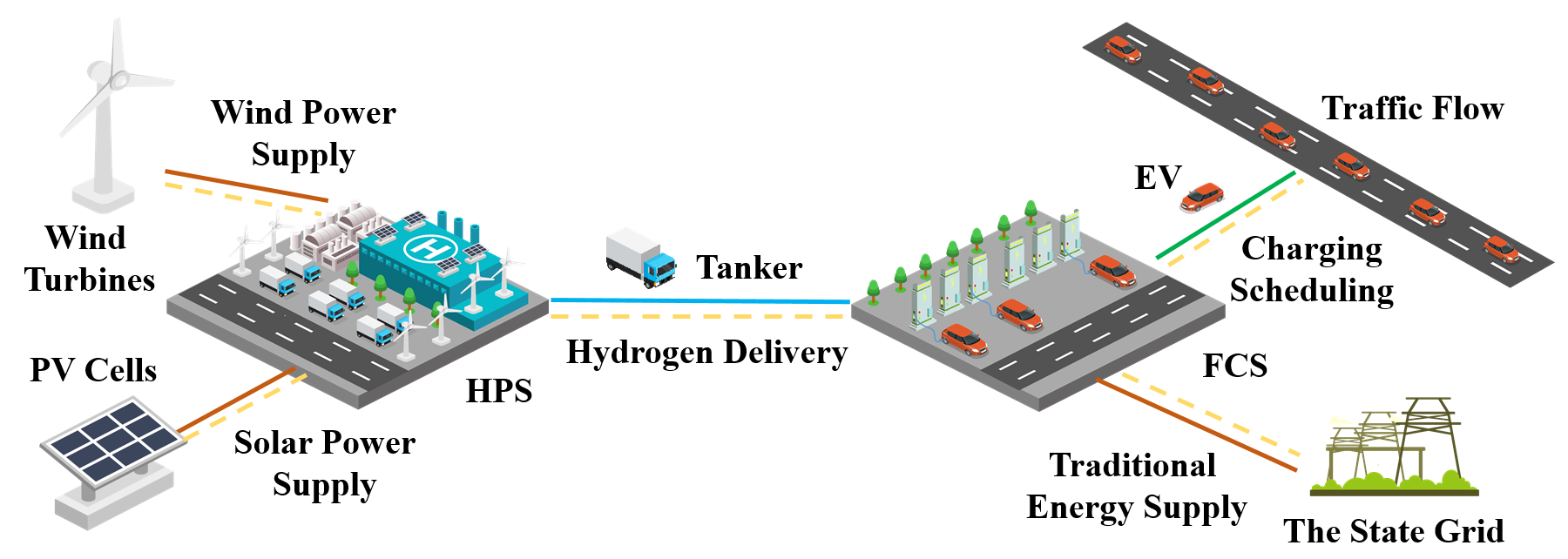}
	%	\captionsetup{margin=20pt,format=hang,justification=justified}
	\caption{The relationship between interconnected elements of the CPES.}
	\label{fig:2}
\end{figure}

The proposed HPS-FCS-EV architecture for joint hydrogen energy schedule and EV coordinated charging is depicted in Fig. \ref{fig:1}. The main stakeholder of the architecture is a company operating several EVs, FCSs and HPSs who wants to minimize the total operating cost by scheduling the hydrogen power dispatch and EV charging location. The company can be a private enterprise such as Uber \cite{uber} and DiDi \cite{didi} that invests in renewable energy and controls the charging plan of EV assets to achieve corporate benefits. It may also represent the municipality which makes efforts to achieve a zero-carbon economy. The detailed relationship between the interconnected elements of the CPES can be found in Fig. \ref{fig:2}. EVs are operated as commercial vehicles to provide passenger services and charged at the FCSs. HPSs and the power grid jointly support the stable operation of FCSs. The hydrogen energy is generated by wind and solar power in decentralized HPSs and transported through tankers. Tankers and EVs share the same transportation network where we distinguish them in Fig. \ref{fig:1} and \ref{fig:2} for clearly explanation. In this paper, we divide the time into equal-length steps and the length of steps is $\Delta$. We make the following assumptions in this paper unless stated otherwise.

$A_1.$ EVs will update some basic (not private) information to the company, such as the charging demand, the state of charge (SoC) and the destination.

$A_2.$ EVs will get fully charged at FCSs and depart at once.

$A_3.$ EVs with passengers will choose higher charging power.

$A_4.$ The charging schedule will not affect the traffic.

Assumption $A_1$ is reasonable since EVs are operated by the company and it is necessary to get some basic (not private) information to make the schedule. Given that the fixed time cost for an EV to charge is usually significant, it tends to get fully charged each time and leave as soon as possible. And for EVs with passengers, the waiting time affects service satisfaction. Thus assumption $A_2$ and $A_3$ hold. Assumptions $A_4$ is reasonable since the number of EVs is tiny for urban traffic. In what follows, we present the models of EV, FCS and HPS in detail.
\vspace{-0.2cm}

\subsection{EV model}
Consider there are $N^{\text{ev}}$ EVs on services. In this paper, we extend the OD flow \cite{zhang2019joint} to describe the EV trajectories under different service states $q_{j,t} = \{0, 1\}$. $q_{j,t} = 1 (0)$ means the EV $j$ is with (no) passengers on board at time $t$.

\nomenclature[I]{$t$}{Indice of time}
\nomenclature[I]{$i$}{Indice of FCSs}
\nomenclature[I]{$j$}{Indice of EVs}
\nomenclature[I]{$k$}{Indice of HPSs}

\nomenclature[P]{$N^{\text{ev}}$}{Number of EVs}
\nomenclature[V]{$q_{j,t}$}{Service state of EV $j$ at time $t$}

\begin{figure}[!htb]
	\centering
	\centering
	\includegraphics[width = 0.45\textwidth]{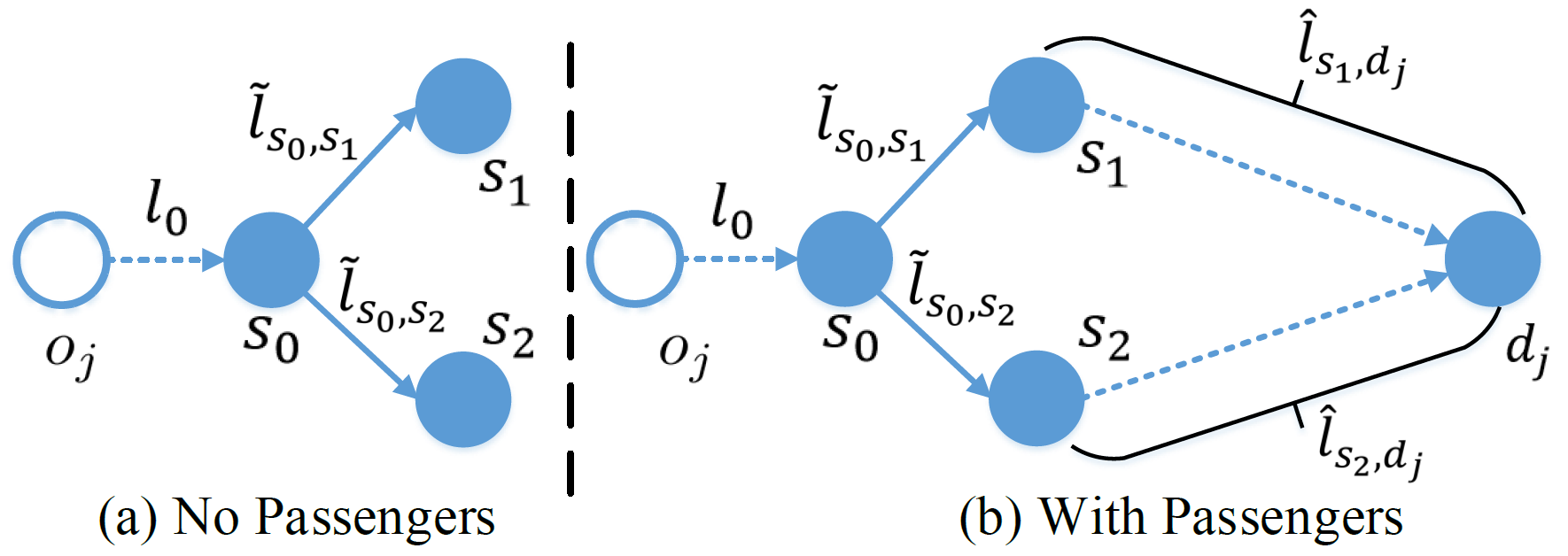}
	\caption{Trajectories of EV $j$ under different service states}
	\label{fig:trajectory}
\end{figure}

Fig. \ref{fig:trajectory} illustrates the typical trajectories of EV $j$ under different service states. A trajectory $\tau_j \in \Omega_t$ of EV $j$, which requests for charging at time $t$, is composed by a set of nodes including an origin node $o_j$ and a destination node $d_j$ (if it has one) and a set of arcs denote the road links between two adjacent nodes. $o_j$ denotes the node where the last recharge was completed. $d_j$ represents the destination of passengers on board. $\Omega_t$ is the set of trajectories of EVs at time $t$.
\nomenclature[V]{$\tau_j$}{Trajectory of EV $j$ at time $t$}
\nomenclature[S]{$\Omega_t$}{Set of trajectories of EVs requesting charging at time $t$.}
\nomenclature[V]{$o_j$}{Origin node of EV $j$}
\nomenclature[V]{$d_j$}{Destination node of EV $j$}
\nomenclature[V]{$SoC_{j,t}$}{SoC of EV $j$ at time $t$}
\nomenclature[P]{$E^{\text{c}}_j$}{Battery capacity of EV $j$ (kWh)}
\nomenclature[V]{$g_{j,t}$}{Charging schedule of EV $j$ at time $t$}

There are several suitable FCSs ($s_1$ and $s_2$ in Fig. 3) nearby with different prices and distances. The charging schedule for EV $j$ can be defined as $g_{j,t} \in \{0,1,...,N^{\text{s}}\}$ where $N^{\text{s}}$ is the number of FCSs. For instance, EV $j$ will be scheduled to be charged at the second FCS if $g_{j,t} = 2$. For EVs do not request charging at time $t$, we set $g_{j,t} = 0$. Different charging schedule will result in different path and distance to the destination node $d_j$. For a no-load EV in Fig. \ref{fig:trajectory}(a), distance and price are the main factors to be considered. However, for an EV with passengers in Fig. \ref{fig:trajectory}(b), different charging schedules will not only affect the charging cost, but also change the path to the destination $d_j$. Thus, the cost function for EV $j$ with $\tau_j$ is as follows,

We assume EV $j$ requests for charging at the node $s_0$ at time $t$. The state-of-charge is $SoC_{j,t}$ which means the charging demand is $(1 - SoC_{j,t}) E^{\text{c}}_j$, where $E^{\text{c}}_j$ is the battery capacity. There are several suitable FCSs ($s_1$ and $s_2$ in Fig. 3) nearby with different prices and distance. The charging schedule for EV $j$ can be defined as $g_{j,t} \in {0,1,...,N^{\text{s}}}$ where $N^{\text{s}}$ is the number of FCSs. For instance, EV $j$ will be scheduled to be charged at the second FCS if $g_{j,t} = 2$. For EVs do not request charging at time $t$, we set $g_{j,t} = 0$. Different charging schedule will result in different path and distance to the destination node $d_j$. For a no-load EV in Fig. \ref{fig:trajectory}(a), distance and price are the main factors to be considered. However, for an EV with passengers in Fig. \ref{fig:trajectory}(b), different charging schedule will not only affect the charging cost, but also change the path to the destination $d_j$. Thus, the cost function for EV $j$ with $\tau_j$ is as follows,
\nomenclature[V]{$C^1_{\tau_j}$}{Total operating cost of EV $j$ with trajectory $\tau_j$ (CNY)}
\nomenclature[V]{$C^{\text{charge}}$}{Charging cost of EV $j$ (CNY)}
\nomenclature[V]{$C^{\text{wait}}$}{Waiting cost of EV $j$ (CNY)}
\nomenclature[V]{$C^{\text{idle}}$}{Idle cost of EV $j$ (CNY)}
\nomenclature[V]{$C^{\text{depre}}$}{Depreciation cost of EV $j$ (CNY)}
\nomenclature[P]{$E^{\text{l}}$}{Power loss per kilometer of EVs (kWh)}
\nomenclature[V]{$\widetilde{\text{l}}_{i,j}$}{Distance between EV $j$ and FCS $i$ (km)}
\nomenclature[V]{$\hat{\text{l}}_{i,j}$}{Distance between FCS $i$ and the destination node of EV $j$ (km)}
\nomenclature[V]{$l_0$}{Distance between origin node and EV $j$ (km)}
\nomenclature[V]{$\beta_{i,t}$}{Charging price of FCS $i$ at time $t$ (CNY)}
\nomenclature[P]{$c^{\text{w}}$}{Per-unit waiting cost (CNY/hr)}
\nomenclature[P]{$c^{\text{i}}$}{Per-unit idle cost (CNY/hr)}
\nomenclature[P]{$c^{\text{d}}$}{Per-unit depreciation cost (CNY/kW)}
\nomenclature[V]{$P_{j,t}$}{Charging power of EV $j$ at time $t$ (kW)}
\nomenclature[P]{$\eta^{\text{c}}$}{Average charging efficiency}
\nomenclature[P]{$v_j$}{Average speed of EV $j$ (km/hr)}
\nomenclature[V]{$E^{\text{pot}}$}{Potential charging demand of EV $j$ (kWh)}
\begin{equation}
C^1_{\tau_j} = C^{\text{charge}} + C^{\text{wait}} + C^{\text{idle}} + C^{\text{depre}}, \ \ \forall \tau_j \in \Omega_t
\end{equation}
where
\begin{equation}
C^{\text{charge}} =  E^{\text{pot}} \beta_{g_{j,t}, t} \tag{1a}
\end{equation}
\begin{equation}
C^{\text{wait}} = q_{j,t} c^{\text{w}}  ( \frac{\widetilde{l}_{g_{j,t},j} + \hat{l}_{g_{j,t},j}}{v_j} + \frac{E^{\text{pot}}}{P_{j,t} \eta^{\text{c}}} )
\tag{1b}
\end{equation}
\begin{equation}
C^{\text{idle}} = (1 - q_{j,t}) c^{\text{i}} \frac{E^{\text{pot}}}{P_{j,t} \eta^{\text{c}}} \tag{1c}
\end{equation}
\begin{equation}
\begin{split}
C^{\text{depre}} = c^{\text{d}} [q_{j,t} (l_0 + \widetilde{l}_{g_{j,t},j} +& \hat{l}_{g_{j,t},j}) + \\
&(1 - q_{j,t}) (l_0 + \widetilde{l}_{g_{j,t},j})]
\end{split}
\tag{1d}
\end{equation}
\begin{equation}
E^{\text{pot}} =  (1 - SoC_{j,t}) E^{\text{c}}_j + E^{\text{l}} \widetilde{l}_{g_{j,t},j} \tag{1e}
\end{equation}

Eq. (1a) describes the charging cost of EV $j$ where $E^{\text{pot}}$ is the potential charging demand and $\beta_{g_{j,t},t}$ denotes the charging price of FCS $g_{j,t}$. $E^{\text{pot}}$ in Eq. (1e) includes the current demand and the power consumption to the FCS where $\widetilde{l}_{g_{j,t},j}$ denotes the distance to FCS $g_{j,t}$ and $E^{\text{l}}$ is the power loss per kilometer. Since waiting time is critical for service evaluation, Eq. (1b) illustrates the waiting cost where $c^{\text{w}}$ denotes the per-unit time cost. The waiting time includes the travel time and charging time where $v_j$ is the speed of EV $j$, $P_{j,t}$ is the charging power, and $\eta^{\text{c}}$ denotes the charging efficiency. For those no-load EVs, charging incurs an unavoidable idle cost given by Eq. (1c) since they cannot operate during that time, where $c^{\text{i}}$ is the per-unit idle cost. Related to the driving distance, the depreciation cost is expressed in Eq. (1d), where $c^{\text{d}}$ denotes the per-unit depreciation cost.

Let $G_t$ denotes the charging schedule matrix, where $G_t(i, j) = 1$ means $g_{j,t} = i$, and it satisfies,
\begin{equation}
\label{con_e1}
\sum_i G_t(i,j) = 0 , \  \forall \tau_j \notin \Omega_t; \ \ \sum_i G_t(i,j) \leq 1 , \  \forall \tau_j \in \Omega_t
\end{equation}

Constraint (\ref{con_e1}) guarantees only EVs requesting for charging will be scheduled to one FCS. Since EVs will not consider the FCSs far away for charging (even if their prices are relatively cheaper), EV $j$ is assumed to only consider FCSs can be reached within $\Delta$, which means,
\begin{equation}
\label{con_e3}
G_t(i,j) \leq R_t(i,j) , \ \ i = 1,2,...,N^{\text{s}}, j = 1, 2, ..., N^{\text{ev}}
\end{equation}
where matrix $R_t$ denotes the available FCS options of EVs, which is defined as follows,
\begin{align}
R_t(i,j) = \left\{ \begin{array}{cc}
1, & \widetilde{l}_{g_{j,t},j} \leq v_j \Delta  \\
0, & \text{otherwise}
\end{array}
\right.
\end{align}

\nomenclature[P]{$N^{\text{s}}$}{Number of FCSs}
\nomenclature[P]{$\Delta$}{Step length of time}
\nomenclature[M]{$R_t$}{Matrix of the charging options of EVs}
\nomenclature[M]{$G_t$}{Matrix of the charging schedule of EVs}
\nomenclature[S]{$\Theta_{i,t}$}{Set of EVs that will depart the FCS in the next time step}
\nomenclature[P]{$a_i^{\text{N}}$}{Total number of charging piles at FCS $i$}
\nomenclature[V]{$a_{i,t}$}{Available number of charging piles at FCS $i$}
\nomenclature[V]{$N_{i,t}^{\text{ev}}$}{Charging number of EVs at FCS $i$}
\nomenclature[V]{$L^{\text{ev}}_{j,t}$}{Remaining charging time of EV $j$ at time $t$}
\nomenclature[P]{$P^{\text{r}}_1$}{Rated charging power of no-load EVs (kW)}
\nomenclature[P]{$P^{\text{r}}_2$}{Rated charging power of EVs with passengers (kW)}
\nomenclature[M]{$H_t$}{Matrix of the dispatched hydrogen energy}
\nomenclature[V]{$d_{i,t}$}{Estimated charging demand of FCS $i$ (kWh)}
\nomenclature[V]{$C^2_{i,t}$}{Operation cost of FCS $i$ (CNY)}
\nomenclature[P]{$c^{\text{m}}$}{Per-unit maintenance cost (CNY/kW)}
\nomenclature[P]{$P^{\text{b,s}}_{i,t}$}{Base load of FCS $i$ (kW)}
\nomenclature[P]{$\beta^{\text{e}}_{t}$}{TOU price of electricity at time $t$ (CNY)}
\subsection{FCS model}
The FCS utilizes the dispatched hydrogen energy from HPSs and electricity from the state grid to charge the EVs parking in the FCS. Let $a_i^{\text{N}}$ denotes the total number of charging piles in FCS $i$ and the number of EVs charging at FCS $i$ is denoted by $N_{i,t}^{\text{ev}}$. Thus, the number of available charging piles $a_{i,t} = a_i^{\text{N}} - N_{i,t}^{\text{ev}}$. Basic information of EVs like $SoC_{j,t}$ will be reported to the FCS. Then we have,
\begin{equation}
\label{fcs1}
SoC_{j,t+1} = SoC_{j,t} + P_{j,t} \eta^{\text{c}} \Delta / E^{\text{c}}_j, \ \ j = 1, 2, ..., N_{i,t}^{\text{ev}}
\end{equation}
\begin{equation}
\label{fcs2}
L^{ev}_{j,t} = (1 - SoC_{j,t}) E^{\text{c}}_j / P_{j,t} \eta^{\text{c}} , \ \ j = 1, 2, ..., N_{i,t}^\text{{ev}}
\end{equation}
\begin{equation}
\label{fcs3}
a_{i,t+1} = a_{i,t} - \sum_j G_t(i,j) + |\Theta_{i,t}| , \ \ i = 1, 2, ..., N^{\text{s}}
\end{equation}
\vspace{-0.2cm}
\begin{equation}
\label{con_s1}
\sum_j G_t(i,j) \leq a_{i,t} \leq a^{\text{N}}_{i} , \ \ i = 1, 2, ..., N^{\text{s}}
\end{equation}

Eq. (\ref{fcs1}) represents the SoC dynamics at time $t$. The remaining charging time $L^{\text{ev}}_{j,t}$ of EV $j$ is given in Eq. (\ref{fcs2}). Thus, the number of available charging piles at time $t+1$ can be calculated via Eq. (\ref{fcs3}) where $\Theta_{i,t} = \{j| L^{\text{ev}}_{j,t+1} = 0\}$ denotes the set of EVs that will depart at time $t+1$. Inequality (\ref{con_s1}) ensures that the charging EVs will not exceed the number of available charging piles. Under Assumption $A_3$, EVs with passengers will choose higher charging power to reduce the charging time, that is, 
\begin{align}
P_{j,t} = \left\{ \begin{array}{cc}
P^{\text{r}}_1, & q_{j,t} = 0 \\
P^{\text{r}}_2, & q_{j,t} = 1
\end{array}
\right.
\end{align}
where $P^{\text{r}}_2 > P^{\text{r}}_1$. The charging price of FCS $i$ is a function of dispatched hydrogen energy $\sum_k H_t(k, i), k = 1, 2, ..., N^{\text{h}}$ from all $N^{{h}}$ HPSs and the charging demand $d_{i,t}$, that is,
\begin{equation}
\label{beta}
\beta_{i,t} = \max ( \frac{P^{\text{b,s}}_{i,t} + d_{i,t} - \sum_k H_t(k, i)}{P^{\text{b,s}}_{i,t} + d_{i,t}} , 0) \times \beta^{\text{e}}_t
\end{equation}
where $\beta^{\text{e}}_t$ is the TOU price of electricity and $P^{\text{b,s}}_{i,t}$ is the base load of the FCS $i$. Since the charging demand is difficult to know accurately in advance, it can be estimated by the historical data. The cost of FCS $i$ only includes the maintenance cost of the charging piles $C^2_{i,t}$, that is,
\begin{equation}
C^2_{i,t} = c^{\text{m}} \sum_{j \in N_{i,t}^{\text{ev}}} G_t(i,j) P_{j,t} , \ \ i = 1, 2, ..., N^{\text{s}}
\end{equation}
where the $c^{\text{m}}$ is the per-unit maintenance cost.  
 
\nomenclature[V]{$P_{k,t}^{\text{w}}$}{Wind power at HPS $k$ (kW)}
\nomenclature[P]{$N^{\text{w}}$}{Number of the wind turbines}
\nomenclature[P]{$P^{\text{c, w}}$}{Capacity of the wind turbine (kW)}
\nomenclature[P]{$v^{\text{r}}$}{Rated speed of the wind turbine (m/s)}
\nomenclature[P]{$v^{\text{ci}}$}{Cut-in speed of the wind turbine (m/s)}
\nomenclature[P]{$v^{\text{co}}$}{Cut-out speed of the wind turbine (m/s)}
\nomenclature[V]{$v_{k,t}$}{Wind speed at HPS $k$ (m/s)}
\nomenclature[P]{$N^{\text{h}}$}{Number of HPSs}

\subsection{HPS model}
In order to ensure the cleanness of hydrogen energy production, wind turbines and photovoltaic cells (PV cells) are considered to produce $H_2$ from water by electrolysis. The wind power generation $P_{k,t}^{\text{w}}$ of HPS $k$ at time $t$ can be calculated using the following equations \cite{sarkar2011mw},
\begin{align}\label{windequ1}
P_{k,t}^{\text{w}} = \left\{ \begin{array}{ccc}
N^{\text{w}} P^{\text{c, w}} & v^{\text{r}} \leq v_{k,t} \leq v^{\text{co}} \\
N^{\text{w}} P^{\text{c, w}}(\frac{v_{k,t}}{v^{\text{r}}})^3, & v^{\text{ci}} \leq v_{k,t} \leq v^{\text{r}}\\
0, & \text{otherwise}\\
\end{array}
\right. 
\end{align}
where $k = 1, 2, ..., N^{\text{h}}$. $v^{\text{ci}}$, $v^{\text{r}}$, $v^{\text{co}}$ and $P^{\text{c, w}}$ are the core parameters of the wind turbine. $N^{\text{w}}$ is the number of wind turbines and $v_{k,t}$ denotes the wind speed at HPS $k$. The power generated by PV cells $P^{\text{PV}}_{k,t}$ can be modeled as\cite{sarkar2011mw},
\begin{equation}
P^{\text{PV}}_{k,t} = P^{\text{c,PV}} f^{\text{PV}} (G^{\text{PV}}_{k,t}/ G^{\text{r,PV}})
\end{equation}
where $P^{\text{c,PV}}$ is the capacity of PV cells. $f^{\text{PV}}$ denotes the efficiency of PV inverters. $G^{\text{PV}}_{k,t}$ and $G^{\text{r, PV}}$ are the current and standard solar radiation intensity, respectively. Thus, the available renewable power of HPS $k$ at time $t$ is,
\nomenclature[V]{$P^{\text{PV}}_{k,t}$}{Solar power at HPS $k$ (kW)}
\nomenclature[P]{$f^{\text{PV}}$}{Efficiency of PV inverters}
\nomenclature[P]{$P^{\text{c,PV}}$}{Capacity of PV cells (kW)}
\nomenclature[V]{$G^{\text{PV}}_{k,t}$}{Solar radiation intensity (W)}
\nomenclature[P]{$G^{\text{r,PV}}$}{Standard solar radiation intensity (W)}

\begin{equation}
P^{\text{a}}_{k,t} = P^{\text{w}}_{k,t} + P^{\text{PV}}_{k,t} - P_{k,t}^{\text{b,H}}
\end{equation}
where $P_{k,t}^{\text{b,H}}$ is the base load of HPSs. The HPS uses alkaline electrolyzer to produce hydrogen, that is \cite{ulleberg2003modeling},
\begin{equation}
n^{\text{H}}_{k,t} = \eta^{\text{F}} I^{\text{ae}}_{k,t} N^{\text{ae}} / 2F = \eta^{\text{F}} P^{\text{a}}_{k,t} N^{\text{ae}} / (2 U^{\text{ae}} F)
\end{equation}
where $n^{\text{H}}_{k,t}$ is the number of moles of hydrogen. $\eta^{\text{F}}$ denotes the production efficiency and $N^{\text{ae}}$ denotes the number of electrolyzers. $I^{\text{ae}}_{k,t}$ and $U^{\text{ae}}$ are the current and voltage of electrolyzers. $F$ denotes the Faraday constant. High-pressure gas cylinders are used for hydrogen storage and the conversion of hydrogen energy to electricity is completed by the full cell, whose models are shown as follows \cite{ulleberg2003modeling},
\begin{equation}
\label{hv}
Q^{\text{H}}_{k,t} = n^{\text{H}}_{k,t} R T^{\text{H}} / p^{\text{H}}
\end{equation}
\begin{equation}
I^{\text{H}}_{k,t} = 2 Q^{\text{H}}_{k,t} F
\end{equation}
\begin{equation}
P^{\text{H}}_{k,t} = I^{\text{H}}_{k,t} U^{\text{H}}_{k} = 2 Q^{\text{H}}_{k,t} F U^{\text{H}}_{k}
\end{equation}
where Eq. (\ref{hv}) is the Clapyron equation. $P^{\text{H}}_{k,t}$ denotes the equivalent hydrogen power at HPS $k$. The total cost of HPSs is shown as,
\begin{equation}
\label{ob1}
C^3_{k,t} = c^{\text{m,w}} P^{\text{w}}_{k,t} + c^{\text{m,PV}} P^{\text{PV}}_{k,t} + c^{\text{t}} \sum_i H_t(k, i)
\end{equation}
where the first two terms represent the maintenance cost of PV cells and turbines. $c^{\text{m,w}}, c^{\text{m,PV}}$ denote the per-unit maintenance cost of turbines and PV cells. The third term denotes the hydrogen delivery cost through tankers, which is related to the dispatch strategy $H_t(k, i)$ and per-unit delivery cost $c^{\text{t}}$. Similar to constraint (\ref{con_e3}), the HPSs can only supply FCSs within a certain distance, which means, 
\begin{align}
\label{con_h1}
H_t(k,i) = \left\{ \begin{array}{cc}
[0, P^{\text{H}}_{k,t}], & L(k,i) = 1  \\
0, & \text{otherwise}
\end{array}
\right.
\end{align}
where matrix $L$ denotes the supply relationship between HPSs and FCSs, that is,
\begin{align}
L(k,i) = \left\{ \begin{array}{cc}
1, & D(k,i) \leq v^{\text{H}} \Delta  \\
0, & \text{otherwise}
\end{array}
\right.
\end{align}
where $D$ is the distance matrix of HPSs and FCSs, $v^{\text{H}}$ is the average speed of tankers. Since the total dispatched power from HPS $k$ can not exceed the hydrogen power, we have,
\begin{equation}
\label{con_h2}
\sum_i H_t(k, i) \leq P^{\text{H}}_{k,t}
\end{equation}

\nomenclature[V]{$P^{\text{a}}_{k,t}$}{Available power at HPS $k$ (kW)}
\nomenclature[P]{$P_{k,t}^{\text{b,H}}$}{Base load of HPS $k$ (kW)}
\nomenclature[V]{$n^{\text{H}}_{k,t}$}{Number of moles of hydrogen}
\nomenclature[P]{$\eta^{\text{F}}$}{Hydrogen production efficiency}
\nomenclature[P]{$N^{\text{ae}}$}{Number of electrolyzers}
\nomenclature[V]{$I^{\text{ae}}_{k,t}$}{Current of electrolyzers (A)}
\nomenclature[P]{$U^{\text{ae}}$}{Rated voltage of electrolyzers (V)}
\nomenclature[P]{$F$}{Faraday constant}
\nomenclature[V]{$Q^{\text{H}}_{k,t}$}{Volume of the high pressure hydrogen ($m^3$)}
\nomenclature[P]{$R$}{Universal gas constant}
\nomenclature[P]{$T^{\text{H}}$}{Standard temperature of gas cylinders (K)}
\nomenclature[P]{$p^{\text{H}}$}{Standard pressure of gas cylinders (MPa)}
\nomenclature[V]{$I^{\text{H}}_{k,t}$}{Current of the full cell (A)}
\nomenclature[P]{$U^{\text{H}}_{k}$}{Rated voltage of the full cell (V)}
\nomenclature[V]{$P^{\text{H}}_{k,t}$}{Hydrogen power at HPS $k$ (kW)}
\nomenclature[V]{$C^3_{k,t}$}{Operation cost of HPS $k$ (CNY)}
\nomenclature[P]{$c^{\text{m,w}}$}{Per-unit maintenance cost of wind turbines (CNY/kW)}
\nomenclature[P]{$c^{\text{m,PV}}$}{Per-unit maintenance cost of PV cells (CNY/kW)}
\nomenclature[P]{$c^{\text{t}}$}{Per-unit deliver cost of hydrogen energy (CNY/kW)}
\nomenclature[M]{$L$}{Matrix of hydrogen supply relationship}
\nomenclature[M]{$D$}{Matrix of distance between HPSs and FCSs}
\nomenclature[P]{$v^{\text{H}}$}{Average speed of tankers (km/hr)}
\nomenclature[V]{$J_t$}{Cost function at time step $t$}

\subsection{Optimization problem}
Based on the models of the HPS-FCS-EV architecture given above, the objective function of the joint problem at time $t$ is,
\begin{equation}
\label{obj1}
J_t = (\sum_{\tau_j \in \Omega_t} C_{\tau_j}^1 + \sum_{i}^{N^{\text{s}}} C^2_{i,t} + \sum_{k}^{N^{\text{h}}} C^3_{k,t} + n^{\text{nc}} \gamma)
\end{equation}
where the last term denotes the penalty. Specifically, $n^{\text{nc}}$ indicates the number of EVs that have failed to get charging services due to the limitation of charging piles, and $\gamma$ is the penalty factor. Thus, the optimization problem of operating cost minimization can be summarized as follows,
\begin{equation}
\begin{split}
\min_{H_t, G_t}& \sum_{t}^{T} J_t\\
\text{s.t.} \ (2)-(10), (12)&-(18), (20)-(22)
\end{split}
\end{equation}

We denote this problem as P1 where it is a MILP. Several commercial optimization solvers such as IBM ILOG CPLEX can be used to solve P1. However, Solving P1 directly will encounter the following difficulties. \emph{First}, P1 assumes that the EV trajectories and renewable energy supply in the future are known in advance, which is unrealistic in the real-time market. Limited information including the current state and the predictable future can be used by the company to make the scheduling control decision. \emph{Second}, the existence of numerously discrete variables, high dimensionality, and great solution spaces, may lead to the explosion of combination which can take hours to solve it \cite{zhang2016optimal}. Heuristic algorithms may speed up this process, but the performance is difficult to guarantee. However, super-time optimization and decision-making with reliable performance is the key to a company's profitability in the real world.

Based on the above considerations, we propose a T-BBG model in the next section which can be solved online and a bi-level receding-horizon optimization method with the performance guarantee is developed.
\begin{figure}[htb]
	\centering
	\includegraphics[width = 0.35\textwidth]{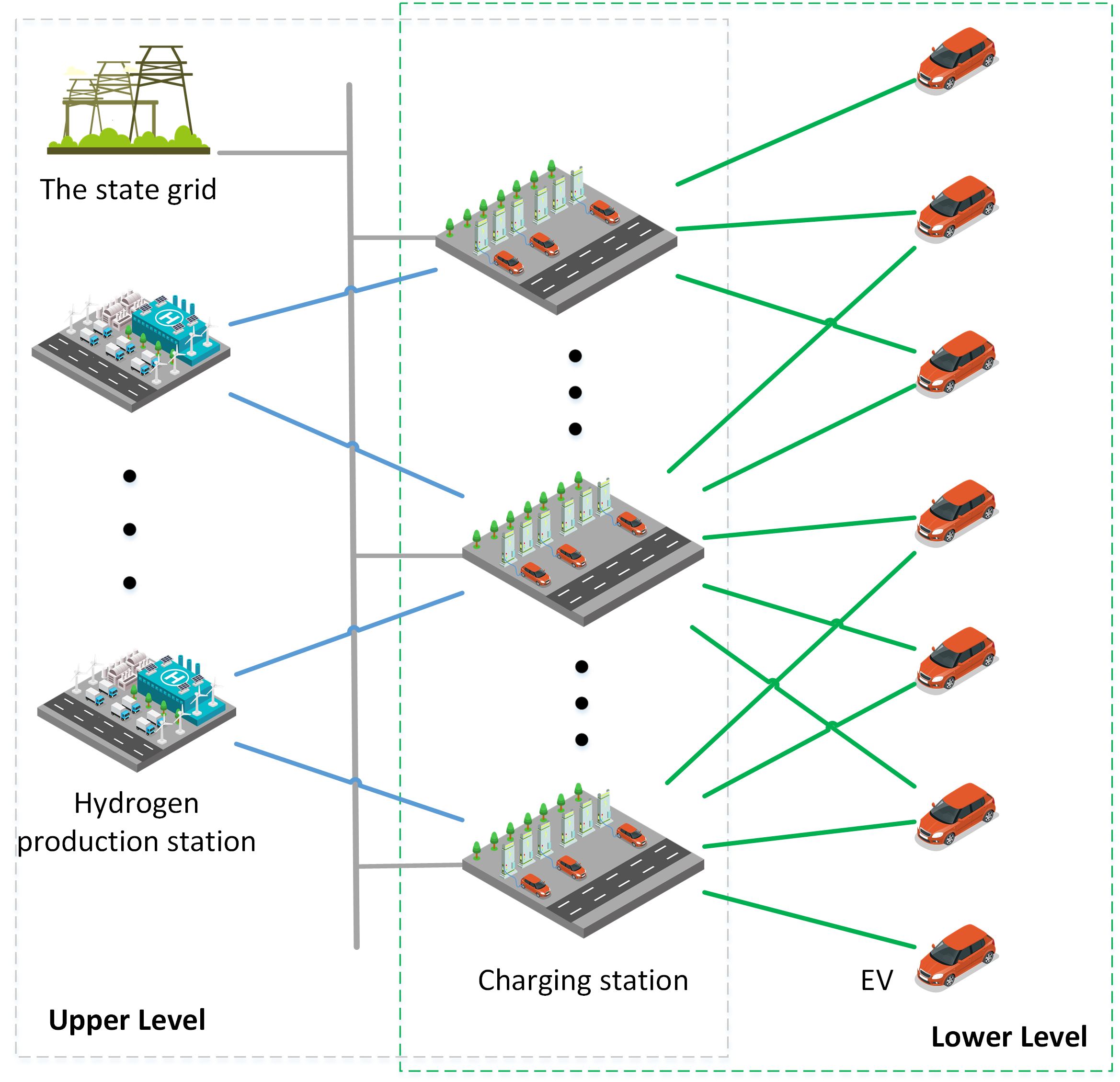}
	\caption{The T-BBG at time step $t$.}
	\label{BBG}
\end{figure}
\vspace{-0.2mm}

\section{Solution Methodology}
\subsection{Time-varying Bi-level Bipartite Graph Model}\label{section_tbbg}
\begin{figure}[htb]
	\centering
	\includegraphics[width = 0.45\textwidth]{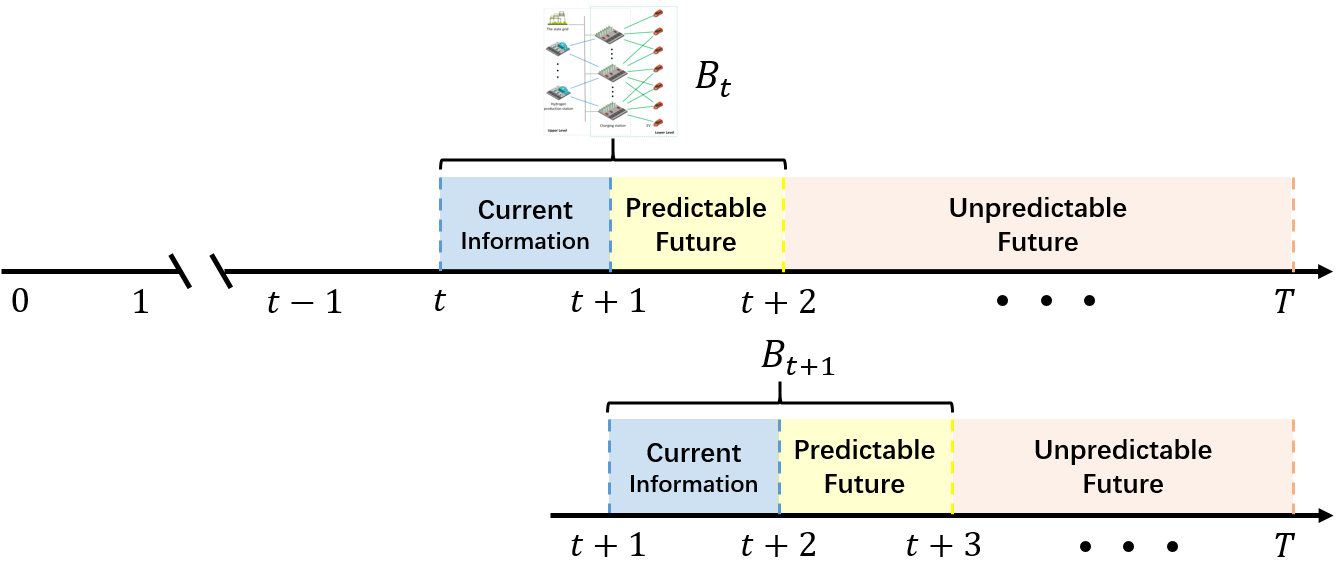}
	\caption{The relationship between the T-BBG and timeline.}
	\vspace{-0.4mm}
	\label{fig:time}
\end{figure}

In the HPS-FCS-EV architecture, the company should make scheduling decisions for the hydrogen energy supply and EV charging demand at each time step. The bipartite graph model effectively represents the supply and demand relationship \cite{bochet2012balancing}. At time $t$, the HPS-FCS-EV architecture can be formulated as a T-BBG $B_t$, which is shown in Fig. \ref{BBG}. The upper level graph (left part) is to dispatch the hydrogen energy to the FCSs, while the lower level graph (right part) denotes the charging location selection problem between EVs and FCSs. Fig. \ref{fig:time} illustrates the relationship between the T-BBG and timeline. Note that $B_t$ is a static slice taken from the timeline when we make decisions and is generated online by scrolling windows. In fact, the nodes, edges, and weights are time-varying which depend on the future supply and demand. Based on (\ref{obj1}), we rewrite the objective function of $B_t$ at time $t$ as,
\begin{equation}
\label{obj2}
J_t = C^{\text{H}} + C^{\text{G,H}} + C^{\text{G}}
\end{equation}
where
\begin{equation}
C^{\text{H}} = \sum_k C^3_{k,t} 
\end{equation}
\begin{equation}
C^{\text{G,H}} = \sum_j C^{\text{charge}}
\end{equation}
\begin{equation}
C^{\text{G}} = \sum_j (C^{\text{wait}} + C^{\text{idle}} + C^{\text{depre}}) + \sum_i C^2_{i,t} + n^{\text{nc}} \gamma
\end{equation}

where $C^{\text{H}}$, $C^{\text{G}}$, and $C^{\text{G,H}}$ denote the cost related to decision variables $H_t$, $G_t$, and both, respectively. Although $H_t$ and $G_t$ affect the objective function together, it can be decoupled and solved iteratively. In what follows, we will elaborate on the problems of the upper and lower levels at time $t$, respectively. 
\nomenclature[V]{$B_t$}{T-BBG model at time step $t$}
\nomenclature[V]{$C^{\text{H/G/G,H}}$}{Cost related to $H_t$/$G_t$/both $H_t$ and $G_t$ (CNY)}

\subsubsection{Upper lower}
Considering any given charging schedule $G_t$ on the lower level (we will discuss this step in detail in \ref{lower}), $C^{\text{G}}$ can be regarded as a constant $c$. $C^{\text{H,G}} = C^{\text{charge}}$ is a piecewise linear function of $H_t$ and $C^{\text{H}} = C^3_{k,t}$ is linear with $H_t$. Thus, the upper level problem can be formulated as a LP, that is,

\begin{equation}
\begin{split}
\min_{H_t} \ & C^{\text{H}} + C^{\text{H,G}} + c\\
\text{s.t.} \ (2)-(10),& (12)-(18), (20)-(22)
\end{split}
\end{equation}

Common LP solvers can be used to optimize the upper level problem and the optimal dispatch strategy can be found.

\subsubsection{Lower level}\label{lower}
\begin{figure}[htb]
	\centering
	\includegraphics[width = 0.45\textwidth]{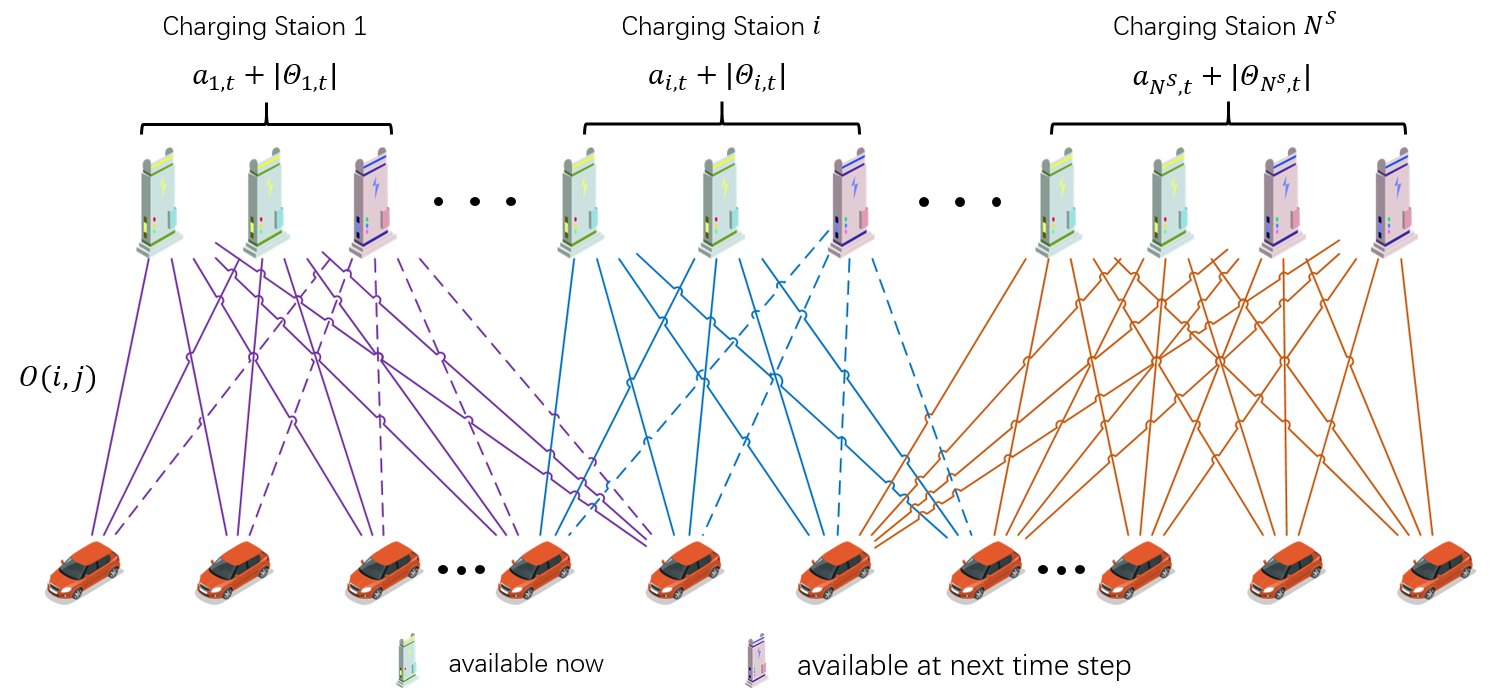}
	\caption{Extended bipartite graph on the lower level.}
	\label{fig:node}
\end{figure}

Similarly, given any the hydrogen energy dispatch $H_t$ ($C^{\text{H}}$ and $\beta_{i,t}$ are constants), the cost related to EV schedule $G_t$ in (\ref{obj2}) is relatively complex. Since the EVs and charging piles in FCSs is a one-to-one matching problem, it can be transferred to a maximum weight matching of an extended bipartite graph by following steps,

Step 1: Since the FCS $i$ can provide $a_{i,t}$ charging services at time $t$, we duplicate $a_{i,t}$ copies of the supply node. Note that there will be at least $|\Theta_{i,t}|$ available charging piles for sure at time $t + 1$, which can give additional options to EVs to wait for one more time step with the extra waiting cost. Therefore, we duplicate $|\Theta_{i,t}|$ copies of the supply node and generate the extended bipartite graph which is shown in Fig. \ref{fig:node}. Thus, the total number of supply nodes (piles) $A_t = \sum_i (a_{i,t} + |\Theta_{i,t}|)$.

\nomenclature[V]{$A_t$}{Number of supply nodes in bipartite graph $B_t$}

Step 2: Let $M_t(i,j)$ denotes the potential total cost of EV $j$ charging at FCS $i$, which can be defined as,
\begin{equation}
M_t(i,j) = C^1_{\tau_j} + c^{\text{m}} P_{j,t} + w(i,j) c^{\text{w}} \Delta
\end{equation}
where $w(i,j)$ is an indicator function indicating whether EV $j$ chooses a pile at time $t + 1$.

Step 3: In order to transform the cost minimization problem into the maximum weight matching problem, we modified the potential total cost $M_t$ to the weight of edges $O_t$ in the bipartite graph, that is,
\begin{equation}\label{ptc1}
O_t(i, j) = \max_{i,j} M_t - M_t(i,j) + 1
\end{equation}

Step 4: The company needs to reduce the operating cost for FCSs and EVs on the premise of ensuring the service rate. In order to meet the charging needs of EVs as much as possible, we set the penalty factor $\gamma$ in Eq. (\ref{obj1}) as,
\begin{equation}\label{gammaeqn}
\begin{split}
\gamma \geq \max (E^{\text{pot}} & \beta^{\text{e}}_t + C^{\text{wait}} + C^{\text{idle}} + C^{\text{depre}} + \\
c^{\text{m}} P_{j,t} + & c^{\text{w}} \Delta) \times \min (A_t, \sum_i N^{\text{ev}}_{i,t})
\end{split}
\end{equation}
then we have the following theorem. 
\begin{theorem}\label{the1}
	To charge the EVs as much as possible is a sufficient condition to get the optimal solution.
\end{theorem}

The proof for Theorem 1 is given in Appendix A. Then this problem at time $t$ is equal to a maximum weight matching of a bipartite graph, and KM algorithm can be used for optimization \cite{west1996}. 
\begin{algorithm}[htb]  
	\caption{Bi-level iteration algorithm for BBG $B_t$}  
	\label{alg:algorithm1}  
	\vspace{.1cm}  
	\begin{algorithmic}[1]   
		\STATE Initialization: choose initial $H_t^0$ and $G_t^0$ randomly, calculate the initial total cost $J_0 = J(H_t^0, H_t^0)$, initialize $\Delta J = \inf$;
		\WHILE{$\Delta J > \epsilon$}
		\STATE Fix the hydrogen energy dispatch $H_t^0$ and modified the lower level as an extended bipartite graph
		\STATE Optimize $G_t^0$ with KM algorithm to get the updated $G_t^1$ and the cost $J_1 = J(H_t^0, G_t^1)$.
		\STATE $G_t^0 = G_t^1$
		\STATE Fix the EV charging schedule $G_t^0$
		\STATE Optimize $H_t^0$ with MILP algorithm to get the updated $H_t^1$ and the cost $J_2 = J(H_t^1, G_t^0)$.
		\STATE $H_t^0 = H_t^1$
		\STATE $\Delta J = |J_2 - J_0|$
		\STATE $J_0 = J_2$
		\ENDWHILE
		\STATE Output: $H_t^0$, $G_t^0$ and $J_0$
	\end{algorithmic} 
	\hrule  
\end{algorithm}

\subsection{Bi-level Iteration Algorithm}\label{section_solution}
Based on the T-BBG model $B_t$ formulated in \ref{section_tbbg}, we first propose a bi-level iteration algorithm to solve the problem at time $t$. It is summarized in Algorithm \ref{alg:algorithm1} where $\epsilon$ is the stopping threshold. Note that when we optimize the schedule of one level, the schedule of another level remains constant as the boundary condition. Based on Theorem \ref{the1}, we can prove the optimality of the proposed algorithm, which is,
\begin{theorem}
	For any hydrogen energy dispatch $H_t$ and EV charging schedule $G_t$ as initialization, Algorithm \ref{alg:algorithm1} can get the optimum.
\end{theorem}

\begin{figure}[htb]
	\centering
	\includegraphics[width = 0.45\textwidth]{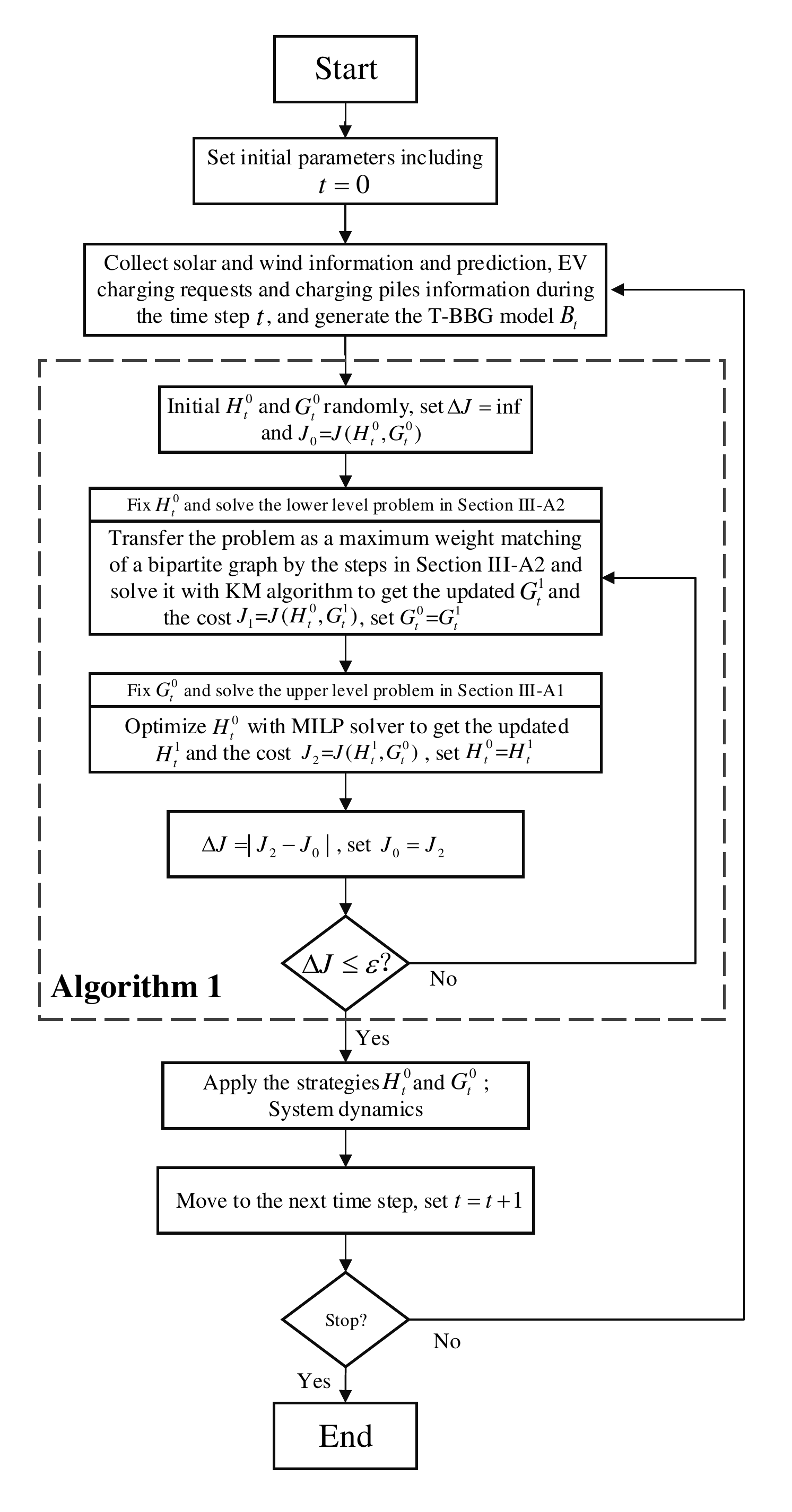}
	\caption{The flowchart of the online control framework.}
	\label{fig:framework}
\end{figure}

The proof for Theorem 2 is given in Appendix B. Considering the optimization of multiple time stages in a day, a receding-horizon online control framework is developed as follows and the detailed flowchart is shown in Fig. \ref{fig:framework}.

Step 1: At time $t = 0$, initialize the system parameters, including the parameters of HPSs, FCSs, and EVs.

Step 2:  Collect the information and prediction of the solar and wind power supply, EV trajectories and charging piles in time step $t$. Generate the T-BBG model $B_t$.

Step 3: Optimize the strategy including hydrogen energy dispatch $H_t$ and EV charging schedule $G_t$ through Algorithm 1.

Step 4: Implement the optimized strategy and the system changes dynamically.

Step 5: Set $t = t + 1$ ($\Delta$ passes in the real time) and jump to Step 2.

\section{Numerical Results}\label{section_case}
\subsection{Case Overview and Parameter Settings}
\begin{figure}[!htb]
	\centering
	\subfigure[Tranportaion network]{\includegraphics[width=0.227\textwidth]{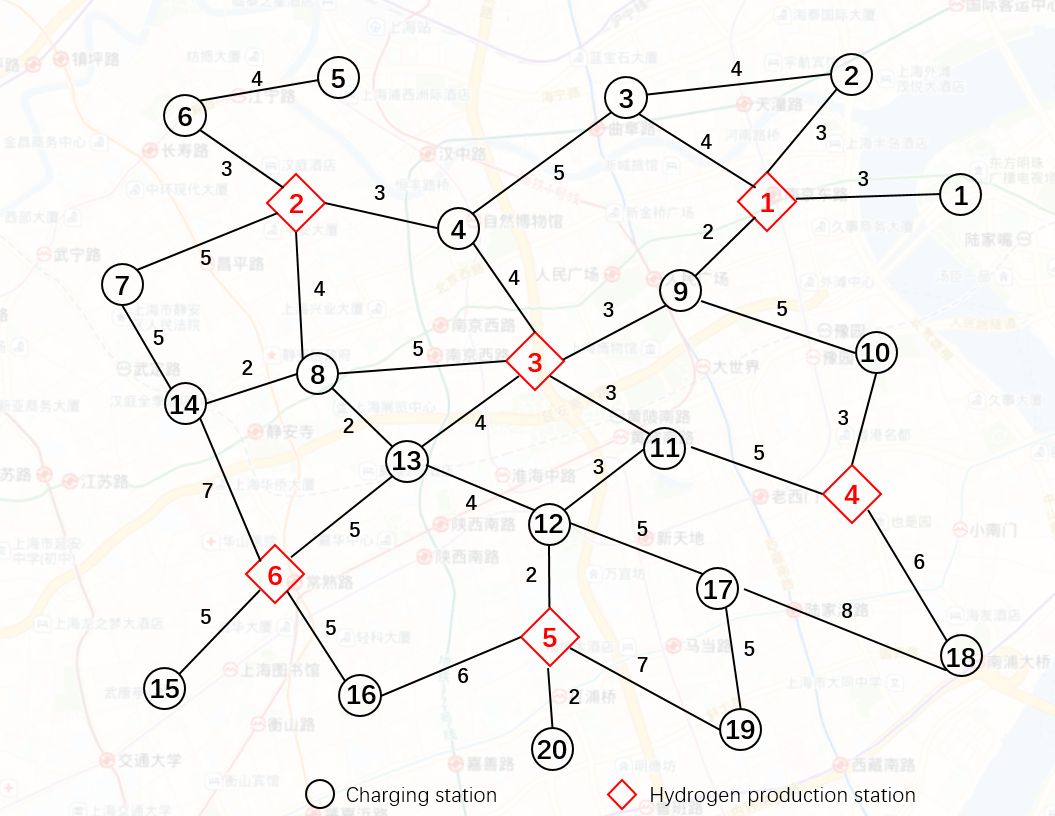}} 
	\subfigure[Heat map of tranportation]{\includegraphics[width=0.22\textwidth]{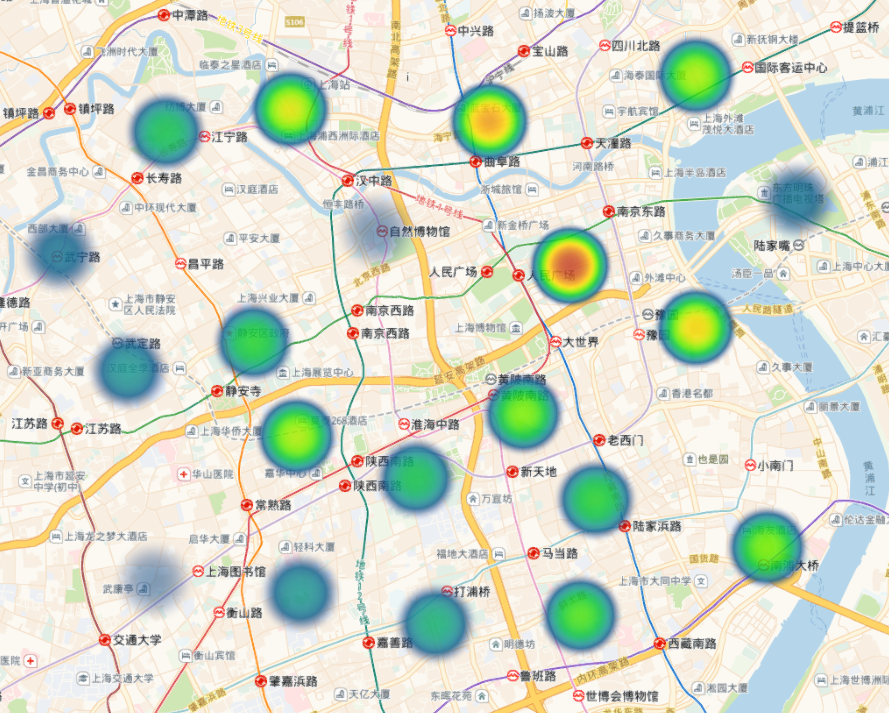}} \\
	\caption{A 26-node HPS-FCS-EV architecture in Shanghai.}
	\label{fig:shanghai}
\end{figure}

In this section, a 26-node transportation network with 20 FCSs and 6 HPSs in Shanghai (see Fig. \ref{fig:shanghai} (a)) is considered to illustrate the proposed architecture. Distance between different nodes is given in the unit of km. For each HPS, one SANY SE13122 wind turbine \cite{SANY} and PV cells with the capacity of 1000 kW are deployed. The real wind speed and solar radiation intensity data in Shanghai collected by the National Meteorological Information Center \cite{wind} are used to generate renewable energy. Detailed parameter settings of the HPSs are shown in Table \ref{table_pa_h}.
\begin{table}[htbp]
	\caption{Parameter settings of HPSs \cite{liuphd2017,Zhuphd2017}}
	\label{table_pa_h}
	\centering
	\begin{tabular}{cc|cc}
		\hline
		\hline
		Parameter & Value & Parameter & Value\\ 
		\hline
		$P^{\text{c,w}}$  	& $2200 \text{kW}$ 	& $v^{\text{r}}$ 	& $12 \text{m/s}$ \\ 
		$v^{\text{co}}$ 	& $22 \text{m/s}$  	& $v^{\text{ci}}$ 	& $2.5 \text{m/s}$\\
		$P^{\text{c,PV}}$ 	& $1000 \text{kW}$ 	& $f^{\text{PV}}$    	& $0.88$\\
		$G^{\text{r,PV}}$ 	& $800 \text{W}$   	& $P^{\text{b,H}}_{k,t}$ & $400 \text{kW}$ \\
		$\eta^{\text{F}}$  	& $0.98$   			& $N^{\text{ae}}$ 		& $8$\\
		$U^{\text{ae}}$ 	& $60 \text{V}$ 	& $F$ 			& $96485.34$\\
		$R$ 		 		& $8.314$  			& $T^{\text{H}}$ 		& $300 \text{K}$ \\ 
		$p^{\text{H}}$ 		& $15 \text{MPa}$ 	& $c^{\text{m,w}}$		& $0.018 \text{CNY/kW}$ \\
		$U^{\text{H}}_k$ 	& $400 \text{V}$  	& $c^{\text{m,PV}}$ 	& $0.018 \text{CNY/kW}$\\
		$v^{\text{H}}$ 	 	& $48 \text{km/h}$	& $c^{\text{t}}$ 		& $0.04 \text{CNY/kW}$\\
		\hline
		\hline
	\end{tabular}
\end{table}
\begin{table}[htbp]
	\caption{Parameter settings of FCSs and EVs \cite{zhang2015integrated, long2017multi}}
	\label{table_pa_ev}
	\centering
	\begin{tabular}{cc|cc}
		\hline
		\hline
		Parameter & Value & Parameter & Value\\ 
		\hline
		$a^{\text{N}}_i$  		& $20$ 			& $P^{\text{b,s}}_{i,t}$ 	& $200 \text{kW}$ \\ 
		$P^{\text{r}}_1$  	& $44\text{kW}$ 		& $P^{\text{r}}_2$  		& $88 \text{kW}$ \\ 
		$\eta^{\text{c}}$	   		& $0.92$ 		& $c^{\text{m}}$	   			& $0.018 \text{CNY/kW}$\\
		$E^{\text{c}}_j$	   	& $75 \text{kWh}$ 		& $E^{\text{l}}$	   		& $0.014 \text{kWh/km}$\\
		$c^{\text{w}}$	   		& $17.2 \text{CNY/h}$ 	& $c^{\text{i}}$	   			& $21 \text{CNY/hr}$\\
		$c^{\text{d}}$	   		& $0.025 \text{CNY/kW}$ & $v_j$	   				& $60 \text{km/h}$\\
		$\gamma$     & $300 \text{CNY}$ & & \\
		\hline
		\hline
	\end{tabular}
\end{table}

There are 20 Mennekes charging piles with two charging modes ($P^{\text{r}}_1 = 44 \text{kW}$ and $P^{\text{r}}_2 = 88 \text{kW}$) at each FCS \cite{rahman2016review}. The TOU price of electricity in \cite{long2017multi} is used. Real commercial taxi data from \cite{evdata} in Shanghai is used to generate time-varying EV trajectories (See Fig. \ref{fig:shanghai} (b)). The company manages 4,000 commercial EVs with about 12,350 charging requests per day. The waiting cost $c^{\text{w}}$ and idle cost $c^{\text{i}}$ are highly connected with passengers' and drivers' income levels which are equal to $70\%$ of the average hourly earnings of non-supervisory employees and taxi drivers in shanghai\cite{stats}. Note that the parameters above are for illustration purposes which should be adjusted in practice. The parameters of the FCSs and EVs are shown in Table \ref{table_pa_ev}.

The schedule time interval $\Delta$ is set to 15 minutes and we consider the control for 24 hours ($T = 96$). The threshold for each time step $\epsilon = 2 \text{CNY}$. We solve this bi-level scheduling problem on a laptop with an 8 core Intel i7-6700HQ processor and 8 GB RAM. To validate the efficacy of our method, $20$ sample paths are generated and the following strategies will be compared:
\begin{enumerate}
	\item MinDistance: Choose the available nearest FCS on the lower level successively and use LP on the upper level.
	\item MinPrice: Choose the available FCS with the cheapest charging price successively on the lower level and use LP on the upper level.
	\item MinCost: Choose the available FCS with the minimum cost function $C^1_{\tau_j}$ successively on the lower level and use LP on the upper level.
	\item NearDis: Dispatch all the hydrogen energy of HPSs to the available nearest FCS on the upper level and use KM algorithm on the lower level.
	\item AveDis: Equally dispatch the hydrogen energy of HPSs to all the available FCSs on the upper level and use KM algorithm on the lower level.
	\item BI-BBG: Algorithm 1 which jointly optimizes the hydrogen energy dispatch and EV charging location selection. 
\end{enumerate}

\subsection{Results Analysis}
\begin{table*}[htbp]
	\vspace{-0.2cm}
	\caption{Optimization results of different strategies}
	\label{table_res}
	\centering
	\resizebox{\textwidth}{!}{
		\begin{tabular}{c|cccccc|c|cc|cc}
			\hline
			\hline
			\multirow{2}*{Strategy} & \multicolumn{6}{c|}{EV cost} & \multicolumn{1}{c|}{FCS cost} & \multicolumn{2}{c|}{HPS cost} & \multirow{2}*{Total cost} &  \multirow{2}*{Std}\\
			\cline{2-10}
			& Charge & Wait & Idle & Depreciation & Penalty & Uncharged & Maintenance & Maintenance & Delivery & &\\ 
			\hline
			MinDistance & 272111.4 & 109152.6 & 114781.9 & 2292.5 & 16500.0 & 55 & 8034.7 & 9648.5 & 10642.6 & 543164.0 & 5419.3\\
			MinPrice 	& 228979.4 & 128371.1 & 120186.7 & 4901.0 & 13200.0 & 44 & 8419.15 & 9648.5 & 10603.1 & 523615.1 & 4772.1\\
			MinCost		& 235991.0 & 116091.4 & 117047.3 & 3117.9 & 12300.0 & 41 & 8204.1 & 9648.5 & 10598.9 & 511812.9 & 4493.6\\
			NearDis & 413347.0 & 100400.3 & 116176.5 & 2482.7 & 2700.0 & 9 & 8134.9 & 9648.5 & 10642.4 & 663532.5& 7255.9\\
			AveDis	& 324982.32 & 108372.0 & 116609.1 & 2612.0 & 2700.0 & 9 & 8161.6 & 9648.5 & 10642.6 & 583728.1& 6437.5\\
			BI-BBG	& 188473.0 & 105784.2 & 117178.0 & 2761.5 & 2700.0 & 9 & 8192.4 & 9648.5 & 10511.7 & 445244.1 & 3598.6\\
			\hline
			\hline
	\end{tabular}}
\end{table*}
\begin{figure*}[!htb]
	\centering
	\subfigure[Number of charged EVs under distance ranking]{\includegraphics[width=0.32\textwidth]{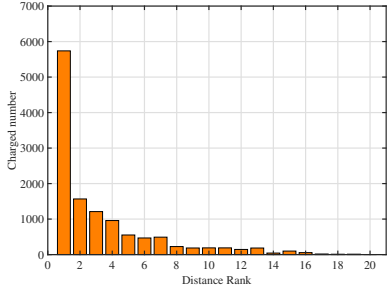}} 
	\subfigure[Number of charged EVs under price ranking]{\includegraphics[width=0.32\textwidth]{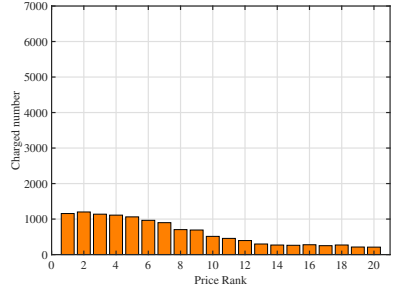}}
	\subfigure[Number of charged EVs under $C^1_{\tau_j}$ ranking]{\includegraphics[width=0.32\textwidth]{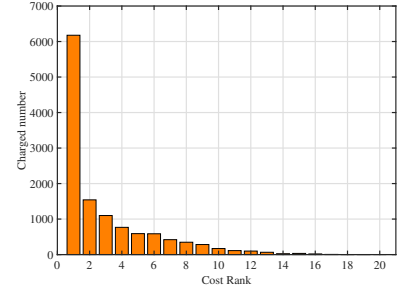}} \\
	\caption{Distribution of charged EVs of BI-BBG strategy under different index orders.}
	\label{fig:rank}
\end{figure*}
The optimization results of different strategies in the case are summarized in Table \ref{table_res}. In general, for a company with 6 HPSs, 20 FCSs and 4,000 EVs, the operating cost for a day is more than 500,000 CNY. And for one charging request, the average charging cost, waiting cost, idle cost, and depreciation cost of EVs are more than 15.26 CNY, 8.57 CNY, 9.48 CNY and 0.19 CNY, respectively. At the same time, it also causes about 0.65 CNY maintenance cost of FCSs, 0.78 CNY maintenance cost and 0.85 CNY delivery cost of HPSs.

As shown in Table \ref{table_res}, the proposed strategy BI-BBG performs best on reducing operating costs as expected. Compared with MinDistance, MinPrice, MinCost, NearDis and AveDis strategies, BI-BBG can reduce the required total cost by about 18\%, 15\%, 13\%, 33\%, 24\%, respectively. For the long-term operation of the company, this cost reduction is significant. Moreover, it demonstrates that the joint optimization of the HPS-FCS-EV architecture (BI-BBG) can achieve better performance than the architectures that only optimize a single problem (MinDistance, MinPrice, MinCost, NearDis and AveDis).

When the NearDis and AveDis strategies are adopted, only the EV charging scheduling problem is optimized. The hydrogen energy dispatch is based on heuristic rules and ignores the dynamic matching of supply and demand sides, which is reflected in the higher charging cost and total cost than other strategies. However, due to the scheduling optimization of EVs, the waiting cost and penalty cost are reduced compared with MinDistance, MinPrice, and MinCost strategies.

When the MinDistance strategy is adopted, EVs tend to choose FCSs closest to the current location. Thus, the idle cost of MinDistance strategy is the least among all strategies, and the waiting cost is relatively less (because the closest FCS may not be in the same direction as the destination, resulting in additional costs). However, since it overlooks other costs (especially the charging cost), its overall cost is relatively high. This cost increase will be more significant in the market dominated by electricity price cost leading to a 22\% increase of the total costs compared with the BI-BBG strategy.

Since the EVs will choose the cheapest charging price under the MinPrice strategy, its charging cost is less than the MinDistance strategy while the waiting and idle cost are higher. Concurrently, it also increases the depreciation cost significantly due to the neglect of the distance factor. 

The Mincost strategy finds a balance between the distance, charging price and other factors. EVs make the charging decisions by considering all possible costs, which makes its total cost is less than the MinDistance and MinPrice strategies. However, However, due to the lack of cooperation between EVs in Mincost strategy, some FCSs with relative price advantages will be quickly occupied, resulting in that the remaining EVs have to choose FCSs with the expensive cost to complete the charging process. The lack of fleet coordination of the above three strategies also increases the uncharged number of EVs, which brings more penalty costs.

By ointly scheduling and coordinating the hydrogen energy and EV charging location, the proposed strategy BI-BBG significantly reduces the overall operating cost of the company at the slight expense of individual optimality, which can be seen from Fig. \ref{fig:rank}. Although most of the EVs are arranged to the FCSs with a shorter distance and lower cost, a small number of EVs are still scheduled to the FCSs with a longer distance and higher cost for the overall performance of the company. Note that in Fig. \ref{fig:rank} (b), the distribution of charged EV number in the FCSs with a lower charging price is almost the same. This is because some FCSs with intensive charging demand are dispatched more hydrogen energy through the upper scheduling. Therefore, their hydrogen energy supply is relatively sufficient and the charging price is basically the same. This shows that the proper schedule of hydrogen energy to achieve the balance of regional matching of supply and demand can bring huge economic benefits, while the delivery cost differences between different schedules may be marginal. It is worth to mention that the BI-BBG strategy not only brings operating cost advantages but also increases the overall service rate (fewer uncharged EV and penalty cost) through the collaborative optimization of two levels of HPS-FCS-EV architecture. This can help the company to spend less on the investment of FCSs and charging piles, which can be a big expense.

\subsection{Sensitivity Analysis}
In this subsection, we analyzed some key parameters of the HPS-FCS-EV architecture, including the pile number, battery capacity, EV speed, and penalty factor. The results can assist the investment decision of the company.
\subsubsection{Pile number}
\begin{figure*}[!htb]
	\centering
	\subfigure[Total cost and service rate]{\includegraphics[width=0.3\textwidth]{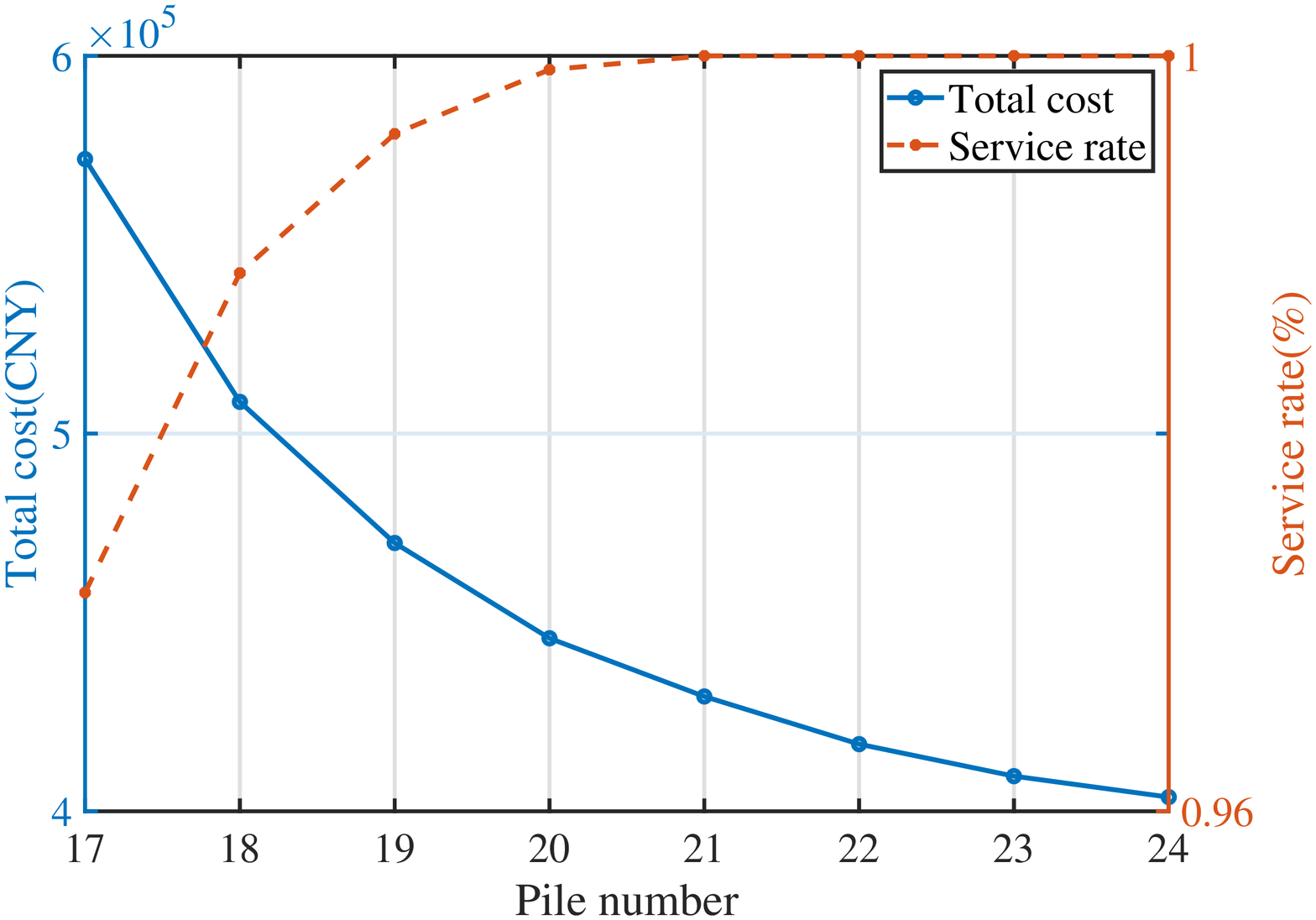}} 
	\subfigure[Charging, waiting and idle cost]{\includegraphics[width=0.3\textwidth]{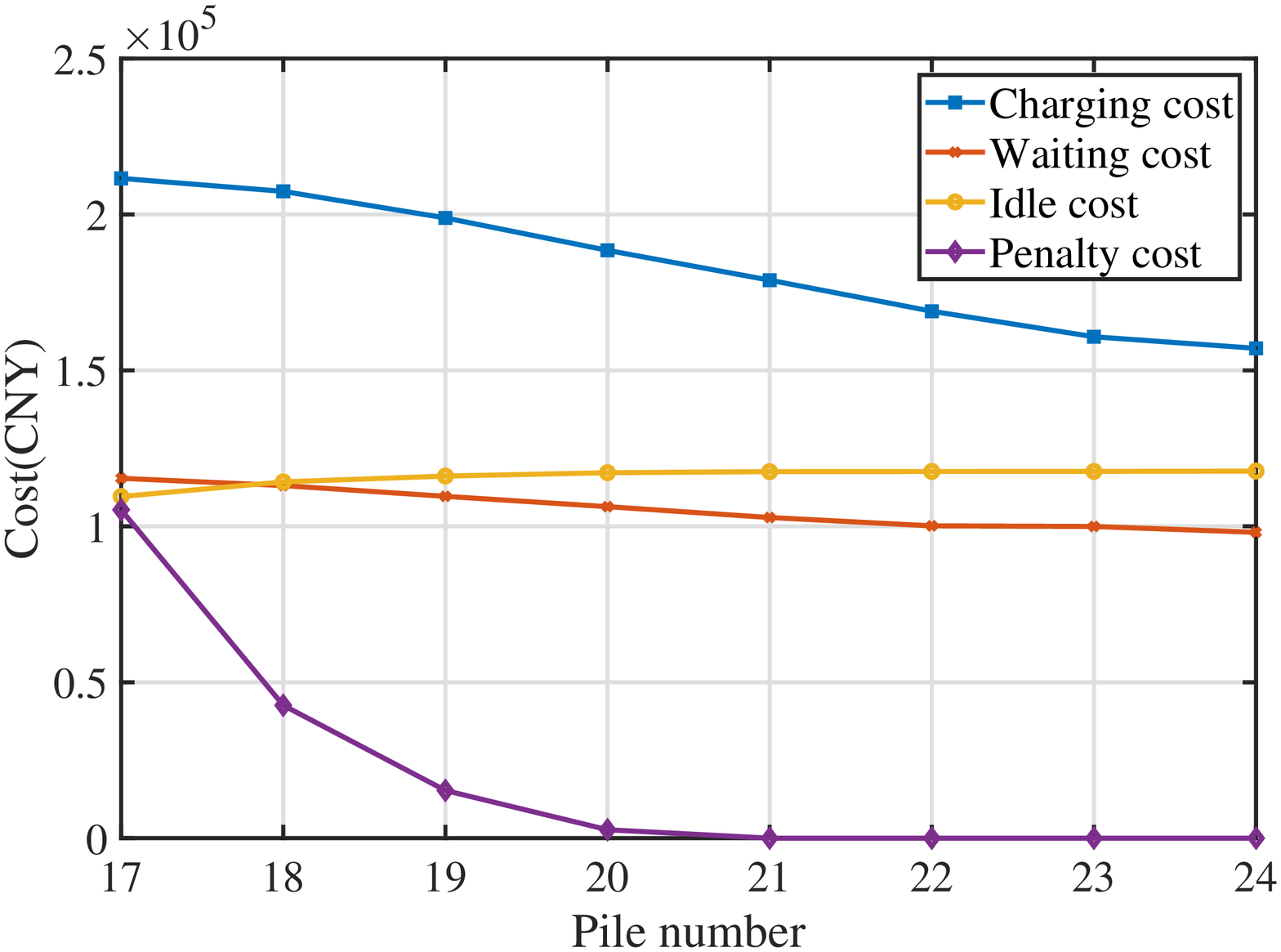}}
	\subfigure[Depreciation, maintenance and delivery cost]{\includegraphics[width=0.3\textwidth]{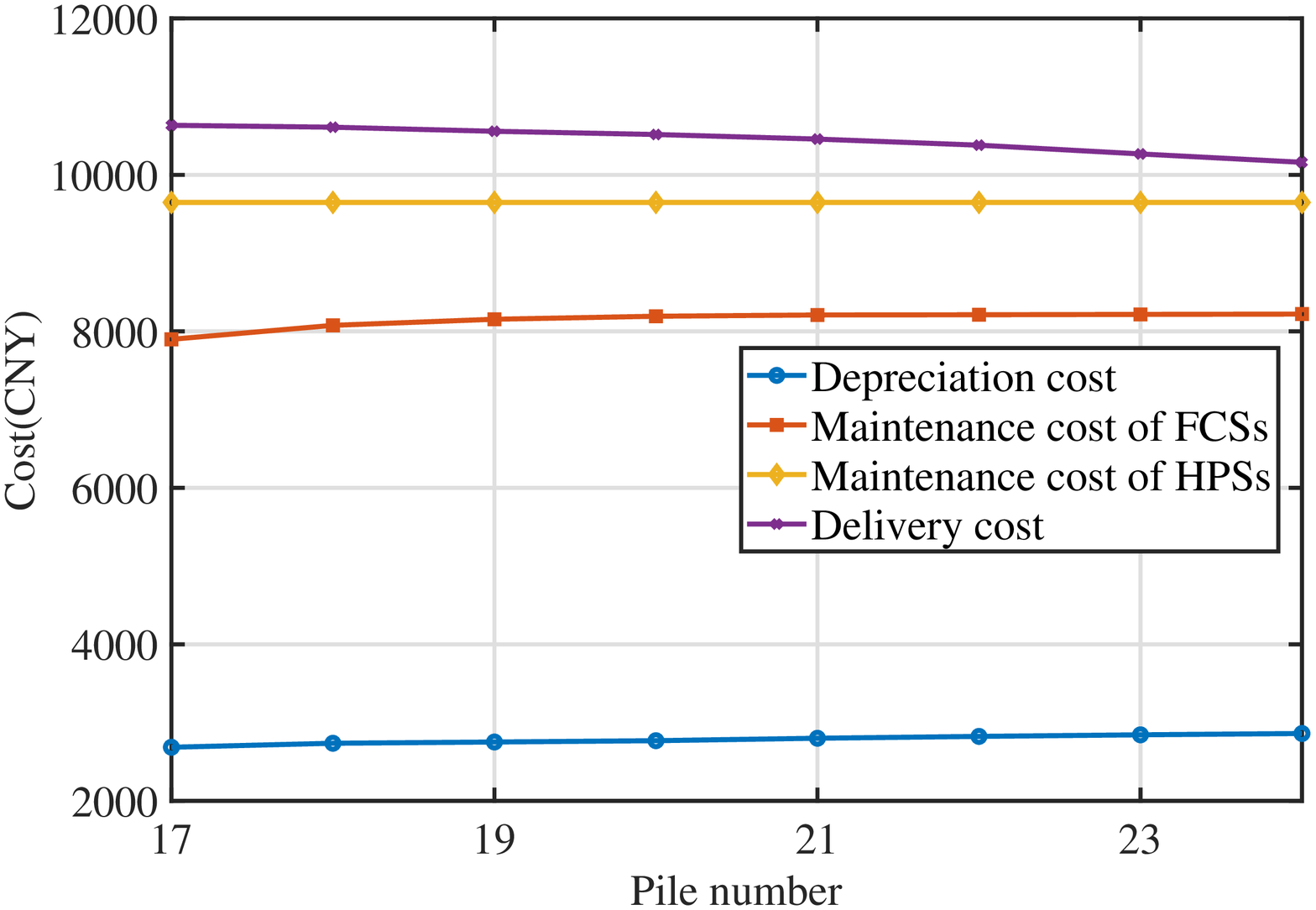}} \\
	\caption{Optimization results with different charging pile number.}
	\label{fig:sen_number}
\end{figure*}

We change the charging pile number in FCSs from $17$ to $24$ to analyzed the impact on the operating cost, and the results are shown in Figure \ref{fig:sen_number}. In general, the total cost is significantly reduced at the cost of additional investment in more charging piles. When the number of charging piles increases, more charging demand can be satisfied in the same time. Thus, the service rate gradually rises to 1 and the penalty cost decreases accordingly. Meanwhile, more EVs can be scheduled to the FCSs with relatively cheaper charging price and shorter distance, resulting in smaller charging cost and waiting cost.

\subsubsection{Battery capacity}
\begin{figure*}[!htb]
	\centering
	\subfigure[Total cost and service rate]{\includegraphics[width=0.3\textwidth]{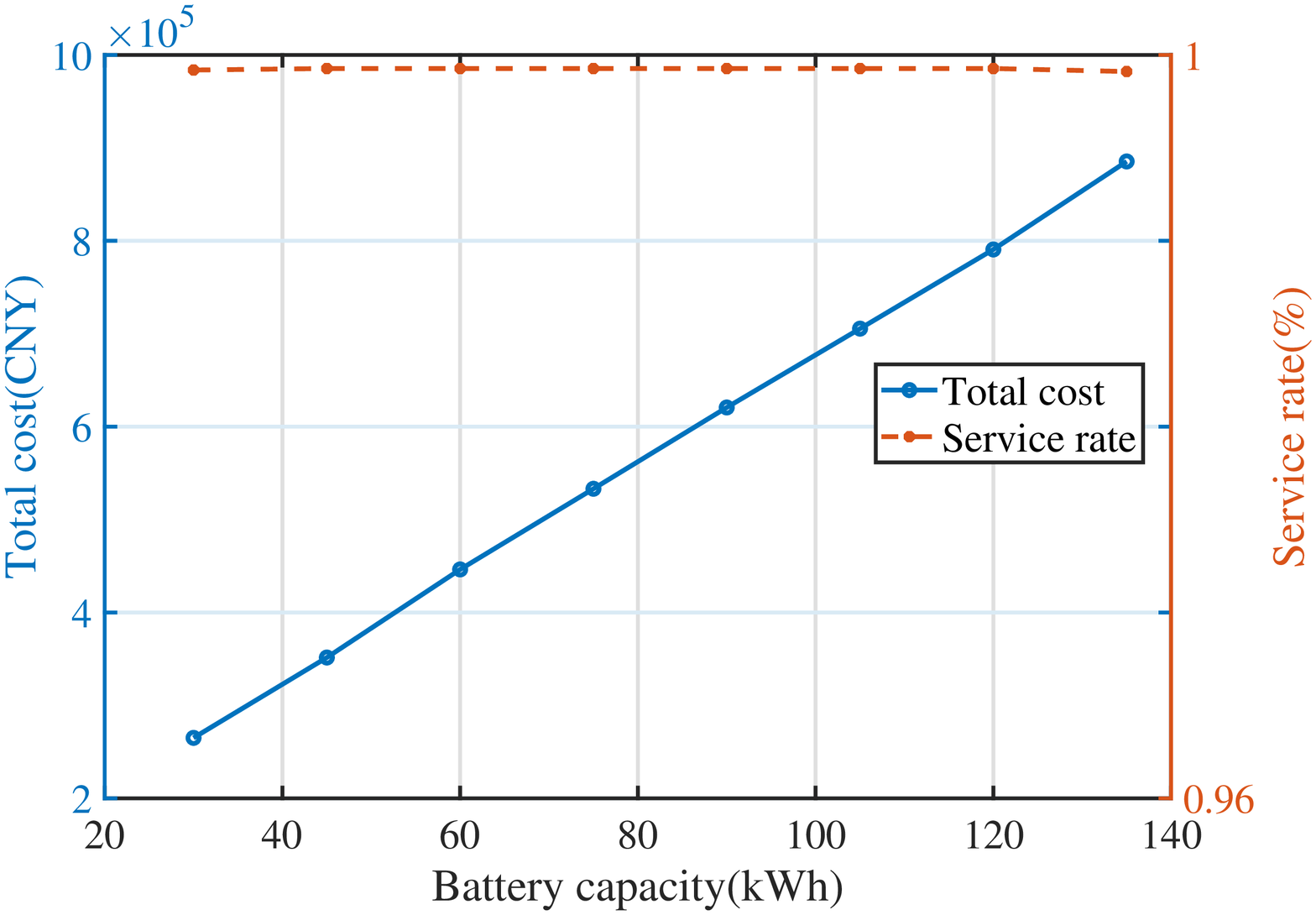}} 
	\subfigure[Charging, waiting and idle cost]{\includegraphics[width=0.3\textwidth]{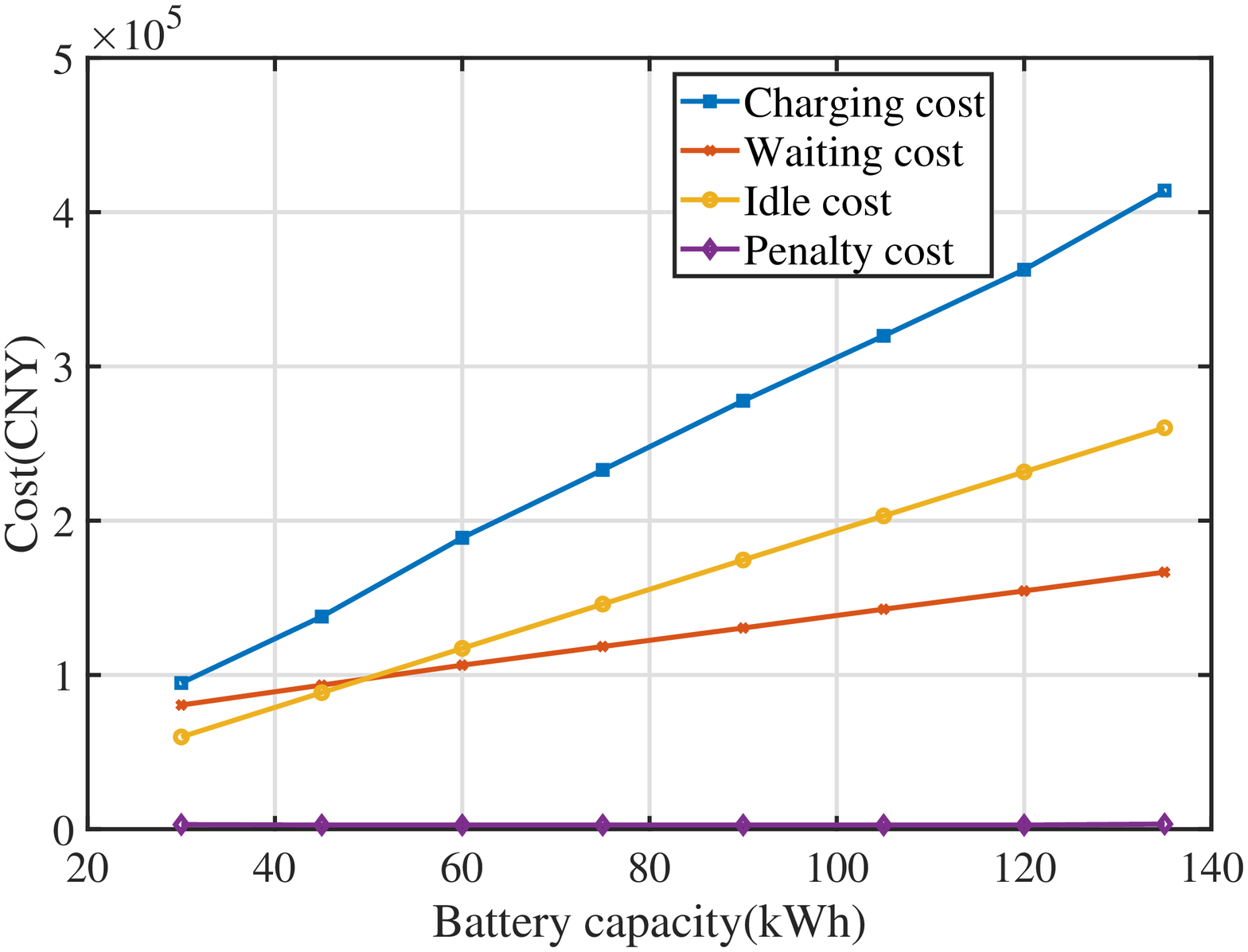}}
	\subfigure[Depreciation, maintenance and delivery cost]{\includegraphics[width=0.3\textwidth]{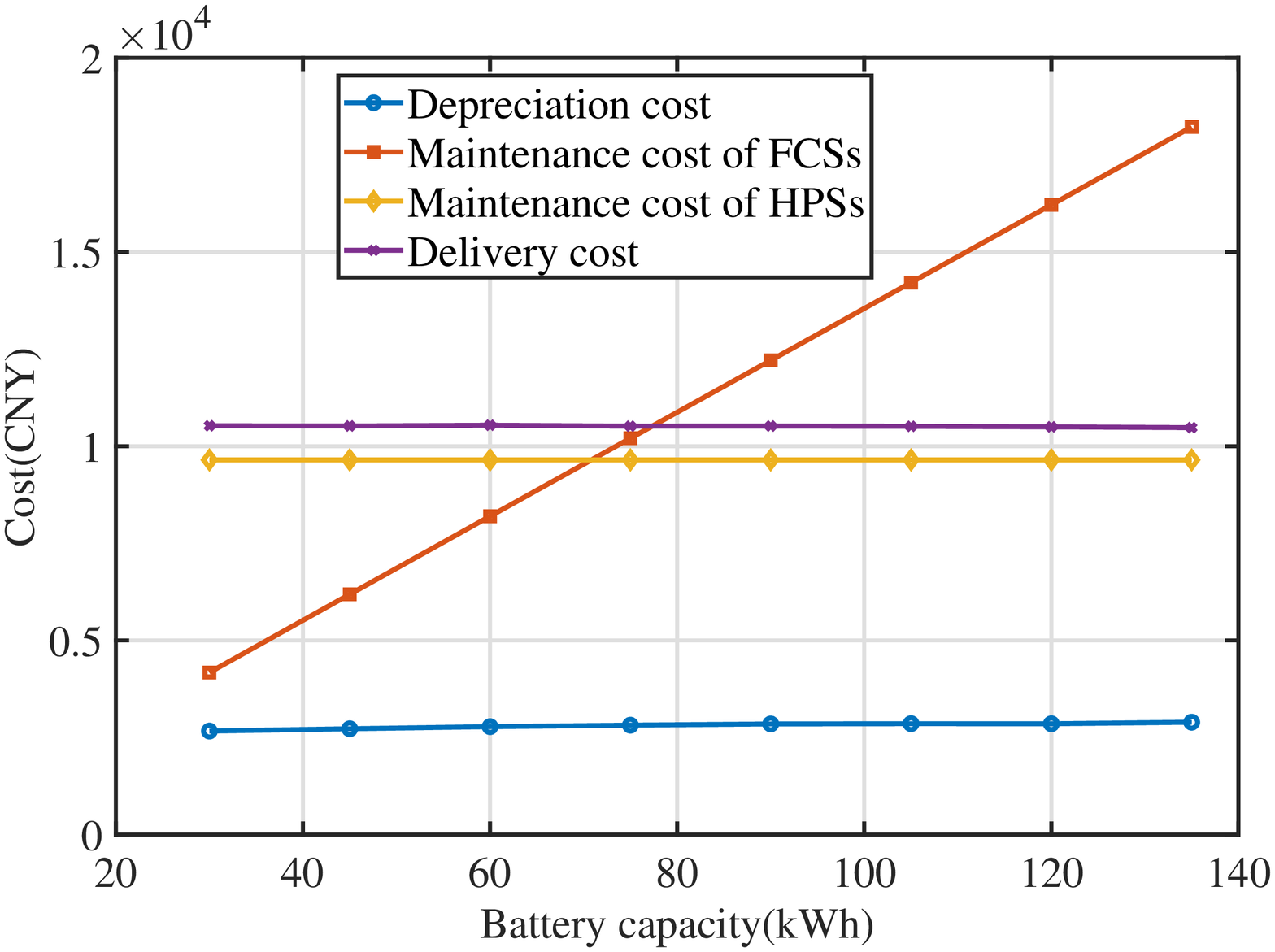}} \\
	\caption{Optimization results with different battery capacity.}
	\label{fig:sen_battery}
\end{figure*}

The impact of battery capacity is analyzed in Fig. \ref{fig:sen_battery}. Assuming that the charging requests are fixed in a day in this setting and change the battery capacity from $20 \text{kWh}$ to $140 \text{kWh}$. The increase of battery capacity will lead to more charging loads and longer charging time. Thus, the charging, waiting, idle cost and maintenance cost of FCSs all increase, while the service rate and other costs remain constant. In fact, larger battery capacity may support a longer driving distance and therefore reduce the charging frequency, which is not discussed in this paper.

\subsubsection{Speed}
\begin{figure*}[!htb]
	\centering
	\subfigure[Total cost and service rate]{\includegraphics[width=0.3\textwidth]{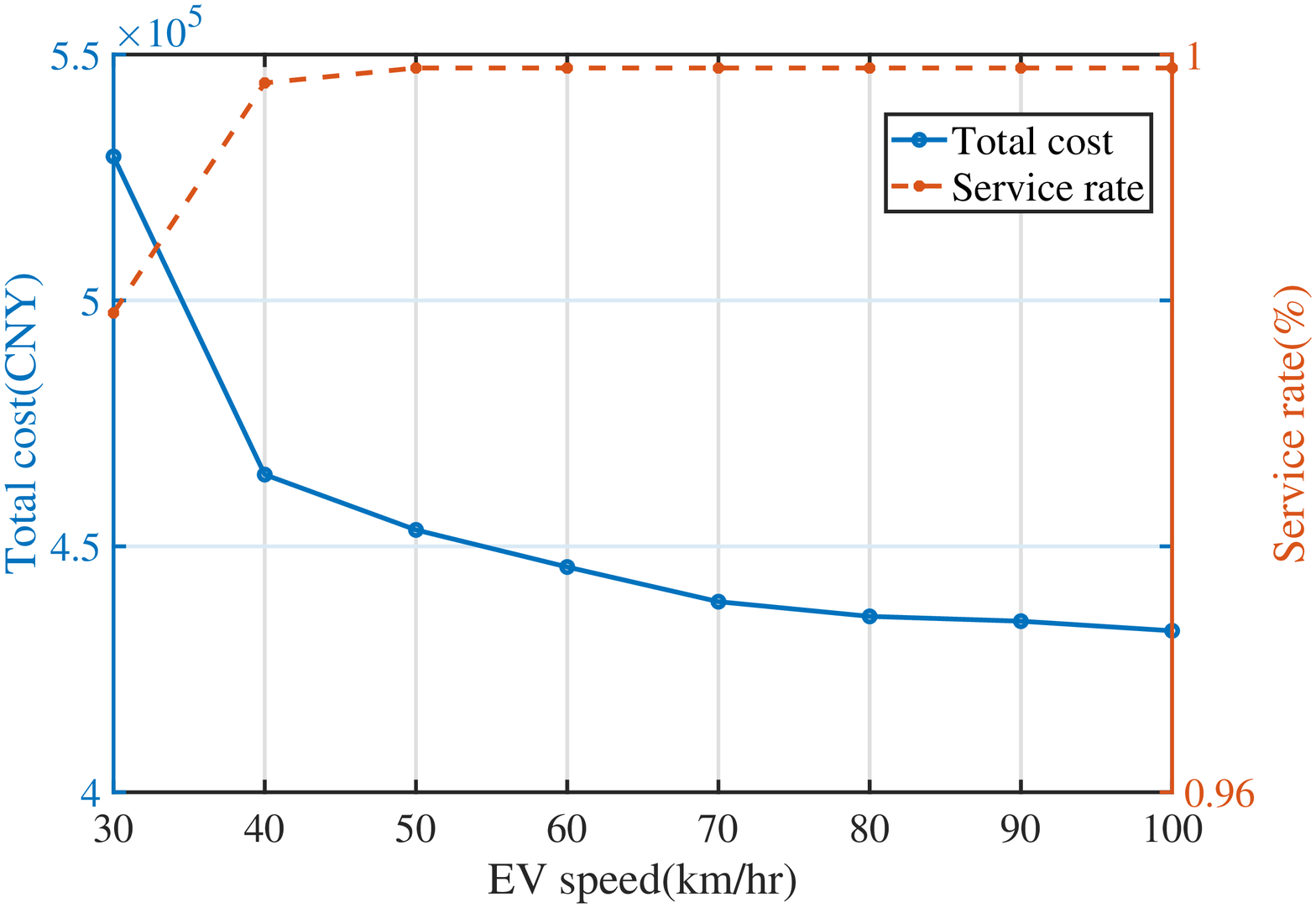}} 
	\subfigure[Charging, waiting and idle cost]{\includegraphics[width=0.3\textwidth]{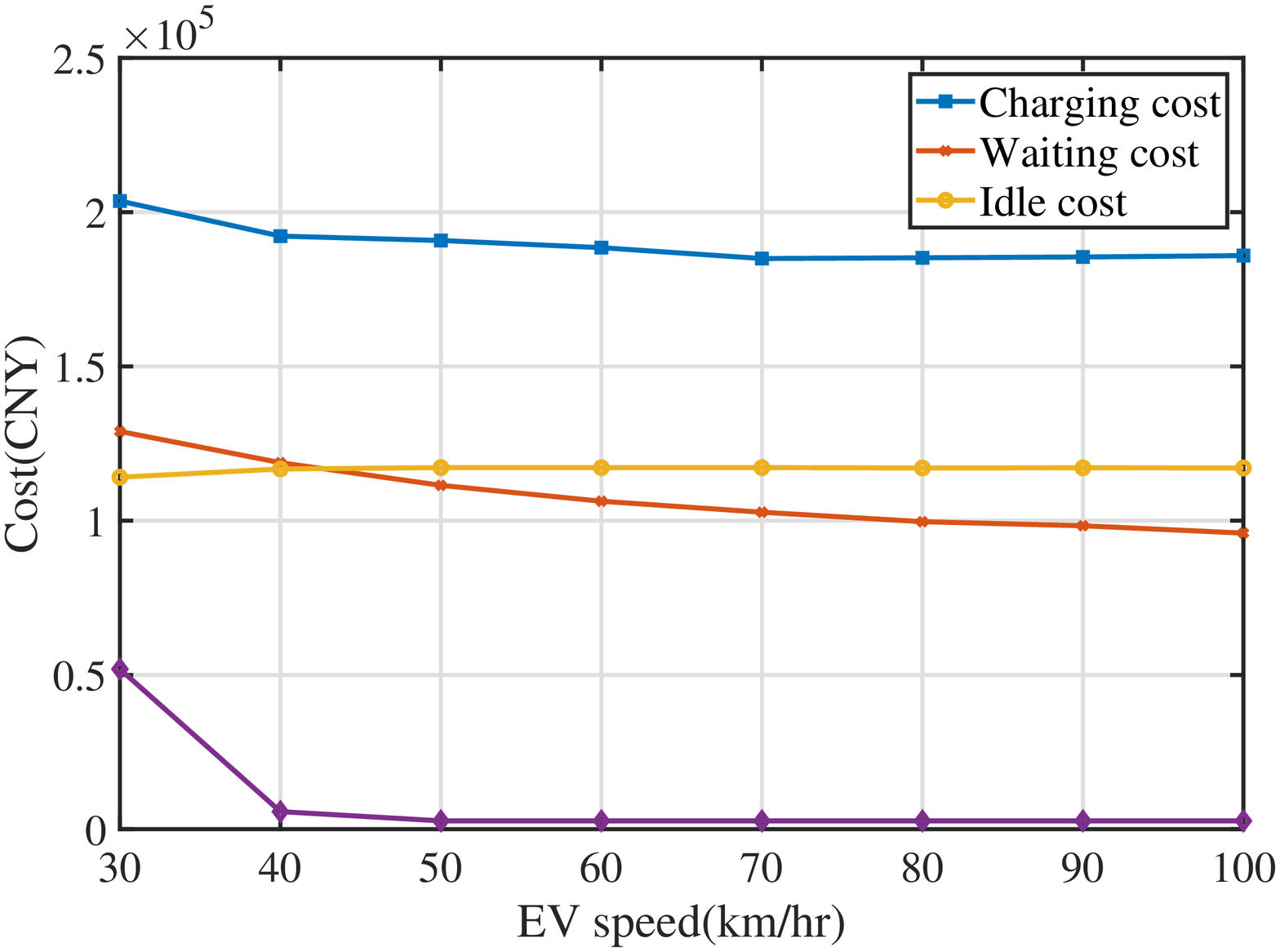}}
	\subfigure[Depreciation, maintenance and delivery cost]{\includegraphics[width=0.3\textwidth]{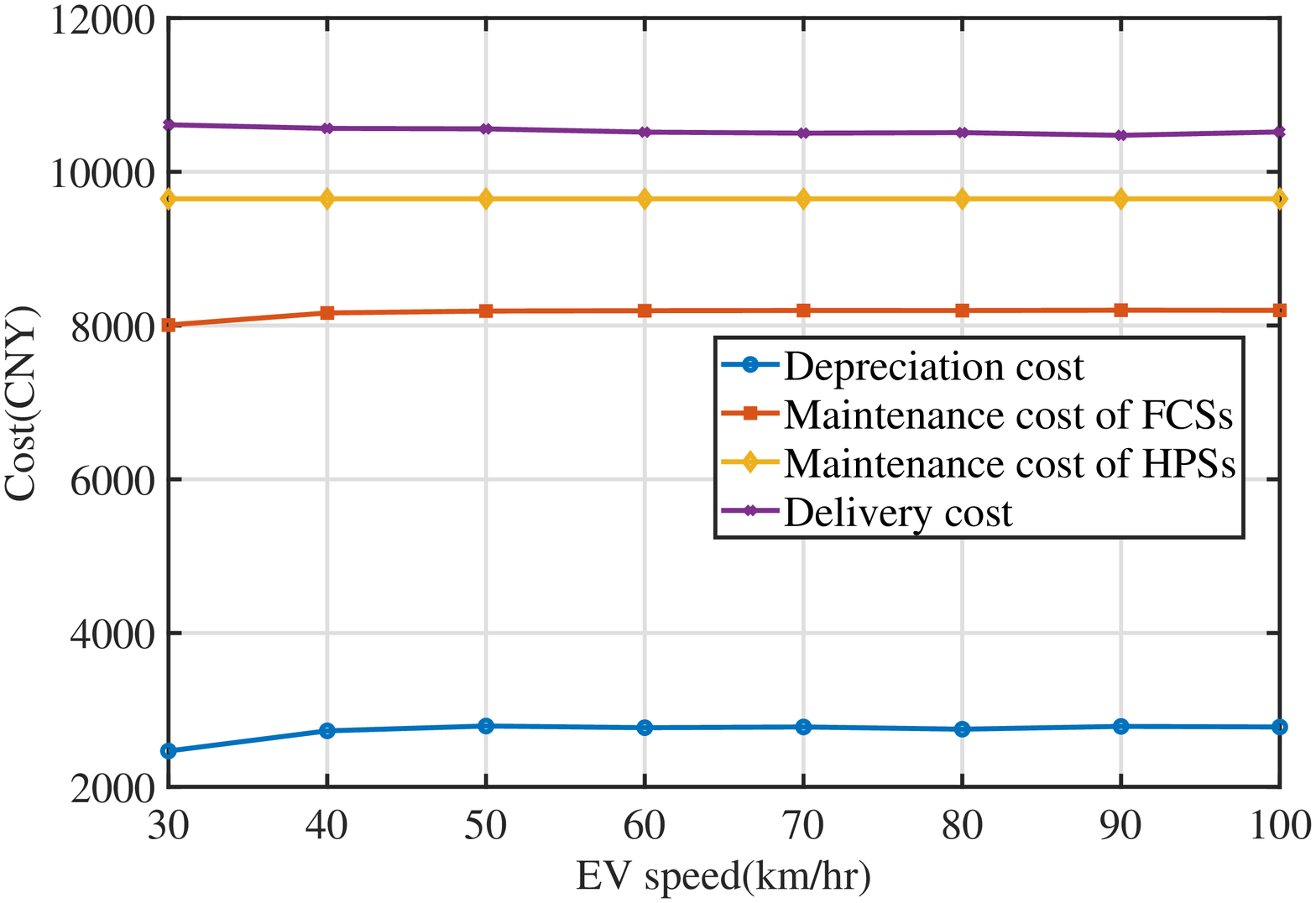}} \\
	\caption{Optimization results with different EV speed.}
	\label{fig:sen_ev_speed}
\end{figure*}

We change the EV speed from $30 \text{km/h}$ to $100 \text{km/h}$ and the impact is evaluated in Fig. \ref{fig:sen_ev_speed}. The total cost falls as the increase of EV speed. As expected, with higher speed, the waiting cost is reduced. Similar with the impact of pile number, EVs with higher speed have more flexibility in scheduling, which means there are more accessible FCSs with lower price and shorter distance. Thus, the charging cost and penalty cost decrease significantly.

\subsubsection{Penalty factor}
\begin{figure*}[!htb]
	\centering
	\subfigure[Total cost and service rate]{\includegraphics[width=0.3\textwidth]{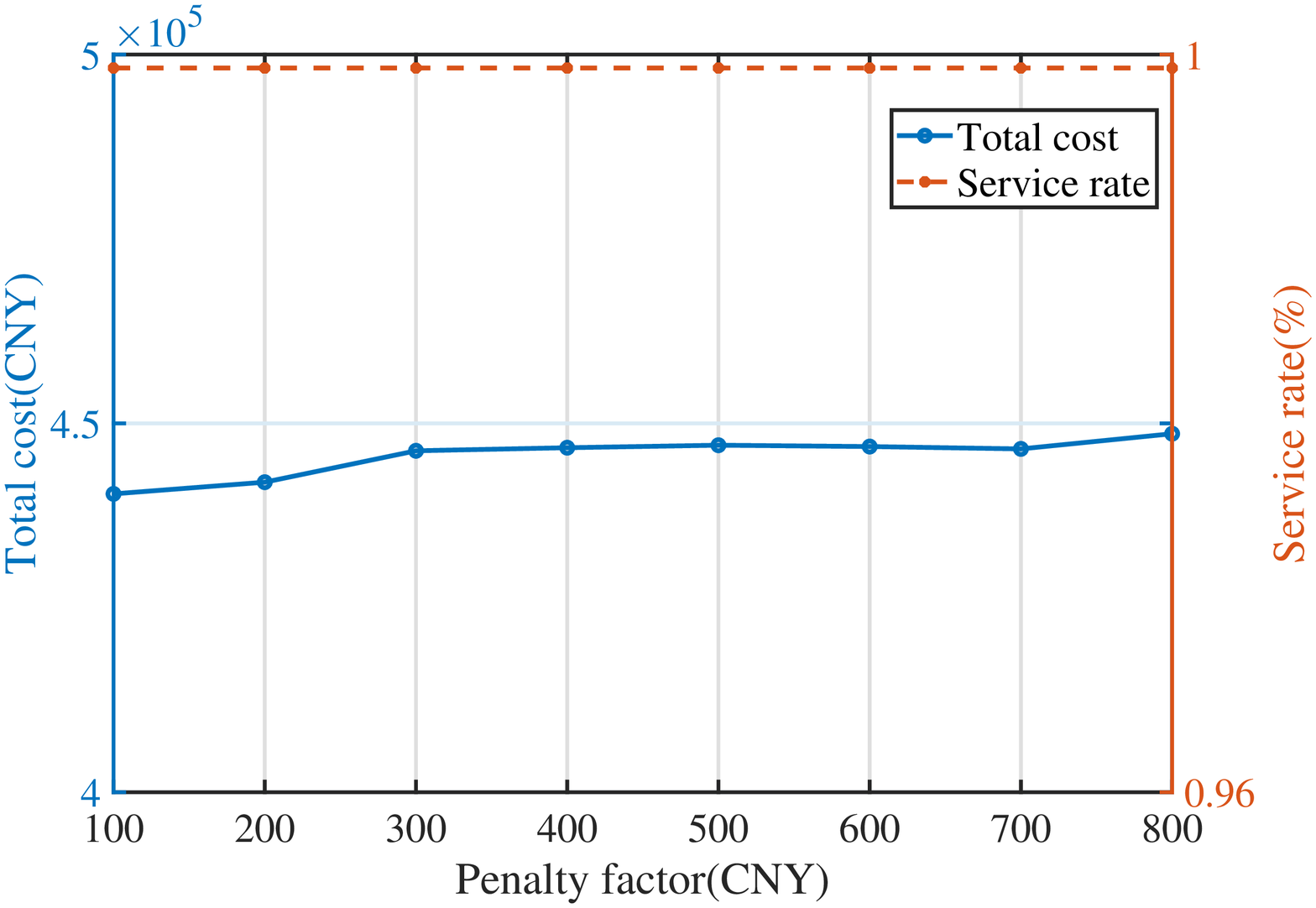}} 
	\subfigure[Charging, waiting and idle cost]{\includegraphics[width=0.3\textwidth]{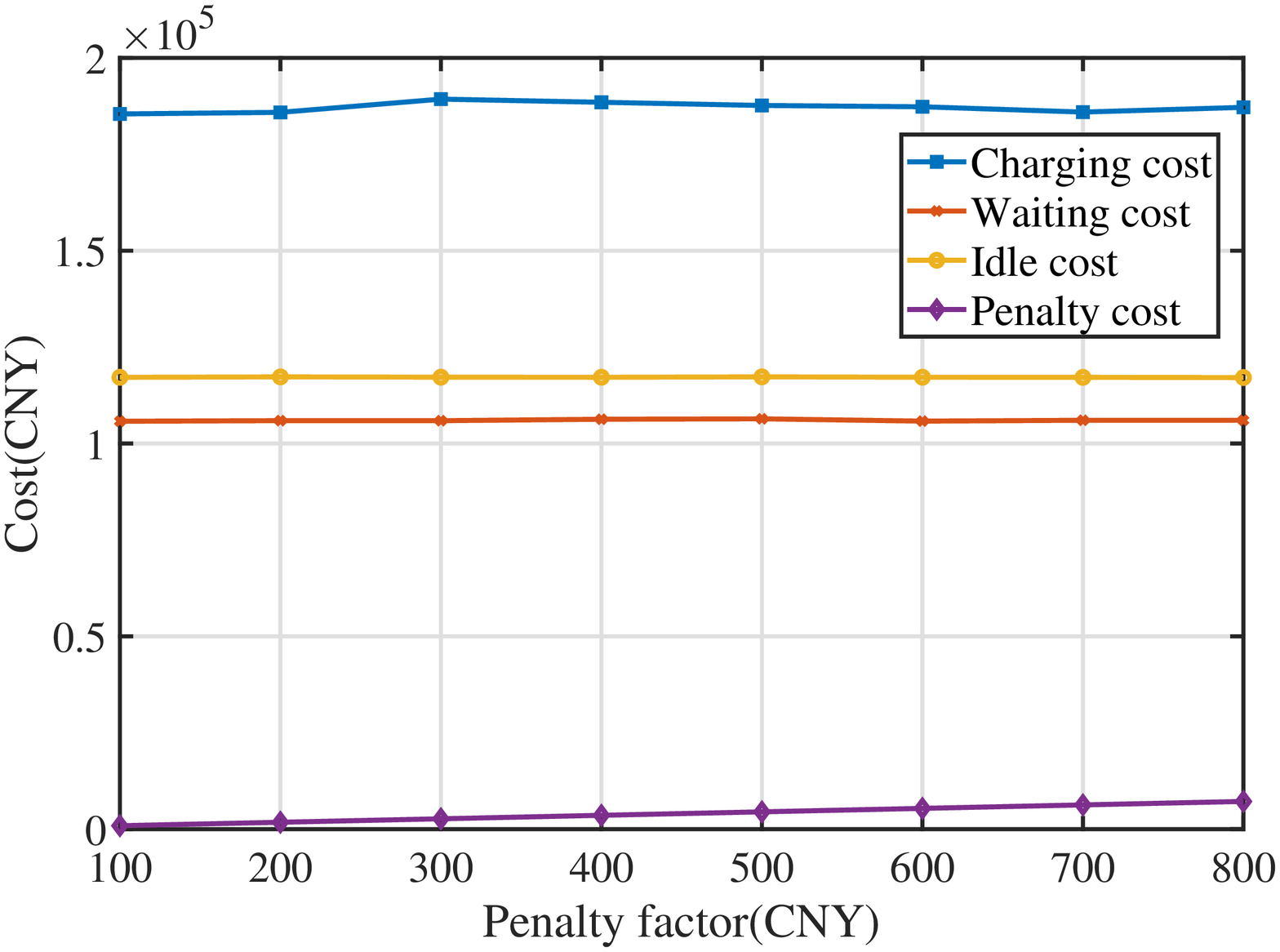}}
	\subfigure[Depreciation, maintenance and delivery cost]{\includegraphics[width=0.3\textwidth]{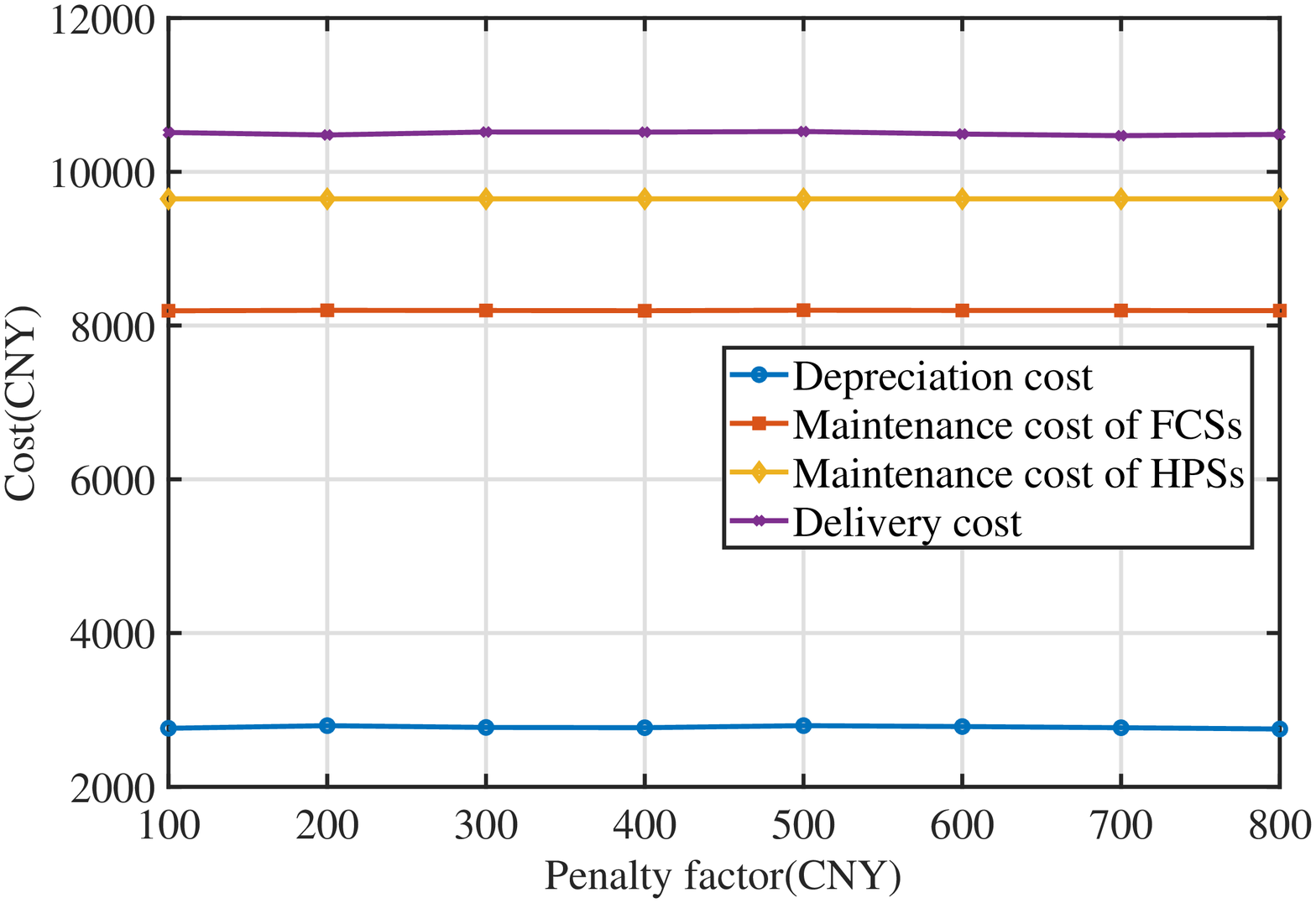}} \\
	\caption{Optimization results with different penalty factor.}
	\label{fig:sen_penalty}
	\vspace{-0.3cm}
\end{figure*}

To illustrate the impact of penalty factor $\gamma$, we conduct the simulation with penalty factors from $100$ to $800$ and the results are shown in Fig. \ref{fig:sen_penalty}. The total cost increases with the increase of penalty factor, which is mainly caused by the increase of penalty cost, while other costs remain almost the same. Since the service rate does not change, the uncharged number of EVs is not affected by the penalty factor. Therefore, the penalty factor actually does not affect the charging scheduling and energy dispatch.

\subsection{Convergence Analysis}
\begin{figure}[htb]
	\centering
	\includegraphics[width = 0.42\textwidth]{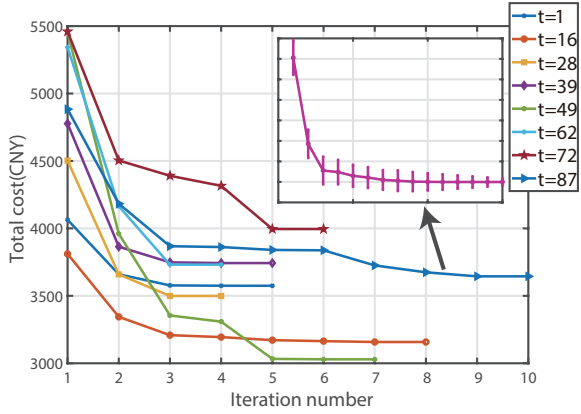}
	\caption{The convergence curve at different time steps.}
	\label{fig:convergence}
\end{figure}

We record the cost change in the iteration process at different time steps. As presented in Fig. \ref{fig:convergence}, the cost change of all time steps shows a monotonic decreasing trend, and finally converges to the minimum. Meanwhile, we use Monte Carlo simulation to randomly generate 300 sample paths at time step 87. The iteration process also converge which can be seen from the subgraph in Fig. \ref{fig:convergence}. The average iteration number of Algorithm \ref{alg:algorithm1} is 4.95 while it cost about 15.4 seconds to get the final scheduling control strategies at one time step. This optimization time is negligible for the online scheduling process, thus the proposed method is competent for the real-time scheduling of a large-scale commercial EV fleet.

\section{Conclusion}\label{section_conclustion}
We proposed a novel HPS-FCS-EV architecture to right-schedule the hydrogen energy dispatch and commercial EV charging location selection jointly. This architecture shows better performance in terms of operating cost savings compared with the ones that consider these two issues separately. A T-BBG model and an efficient bi-level iterative algorithm for real-time scheduling control were proposed and the performance was guaranteed by theoretical analysis and numerical examples. Numerical experiments validated that the proposed method can reduce the operating cost while increasing the service rate. Various parameters' impact was analyzed to help the company make decisions more wisely. 

In this paper, we assumed that EVs requesting charging at the same time will be coordinated synchronously, which is a mild constraint when the interval of time steps is relatively small. However, the charging demand is updated all the time in the real-time operation, which will make our control strategy become conservative and sub-optimal. Our future work will relax this assumption and consider the asynchronous scheduling for EVs. Meanwhile, the main consideration of this paper is the minimization of operating costs, but not the maximization of revenue. In fact, when the marginal utility is positive, an appropriate increase in operating costs can bring greater profits. This will also be our future focus.

% or


\appendices
\section{Proof of Theorem 1}
\proof
Theorem \ref{the1} is equivalent to proving that no matter what values $H_t$ and $G_t$ take, as long as the number of charged EVs satisfies $n_1 < n_2$, there will be $J_{n_1} > J_{n_2}$. $J_{n_1}$ denotes the total cost when the charged number is $n_1$. From (\ref{beta}), we know that regardless of $H_t$, 
\begin{equation}
\begin{split}
\gamma \geq max(E^{\text{pot}} & \beta_{i,t} + C^{\text{wait}} + C^{\text{idle}} + C^{\text{depre}} + \\
c^{\text{m}} P_{j,t} + &c^{\text{w}} \delta_t) \times \min (A_t, \sum_i N^{\text{ev}}_{i,t})\\
= \max_{i,j} M_t&(i,j) \times \min(A_t, \sum_i N^{\text{ev}}_{i,t}))
\end{split}
\end{equation}

Then we prove the theorem by induction. When the first EV is arranged (for example, $G_1(i,j) = 1$), we have that, 
\begin{equation}
J_1 - J_0 = (M(i, j) + (\sum_i N^{\text{ev}}_{i,t} - 1) \gamma) - \sum_i N^{\text{ev}}_{i,t} \gamma <0
\end{equation}

Suppose when the charged number is $n$, it means there are $n$ links in the bipartite graph. When it increases to $n+1$, there must be an augmented chain \cite{johnson1962connectivity}. Consider the longest chain which has a set of $n$ links defined as $s^{\text{c}}$ to be cut, and a set of $n+1$ new links defined as $s^{\text{g}}$ will be generated. $n + 1$ satisfies $n + 1 \leq \min(A_t, \sum_i N^{\text{ev}}_{i,t})$, which means the maximum number of charged EVs is limited by the charging demand and the number of supply nodes. Without loss of generality, we have,
\begin{equation}
\begin{split}
&J_{n+1} - J_{n} = \sum_{i \in s^{\text{g}}} M_{i} - \sum_{j \in s^{\text{c}}} M_{j}  - \gamma\\
< & (n+1 - \min(A_t, \sum_i N^{\text{ev}}_{i,t}))  \max_{i,j} M(i,j) - \sum_{j \in s^{\text{c}}} M_{j} \\
< & 0 
\end{split}
\end{equation}

To sum up, the proof is complete.
\endproof

\section{Proof of Theorem 2}
\proof 
Let $J_0 =J(H_t^0, G_t^0)$. When $H_t^0$ is fixed, the problem is equal to a maximum weight matching of bipartite graph and KM algorithm is applied to optimize it. Define the updated charging schedule as $G_t^1$ and $J_1 = J(H_t^0, G_t^1)$. Since the KM algorithm can find the maximum matching of the bipartite graph, thus $\sum_{i,j} G_t^1(i,j) \geq \sum_{i,j} G_t^0(i,j)$, and, 
\begin{enumerate}
	\item If $\sum_{i,j} G_t^1(i,j) > \sum_{i,j} G_t^0(i,j)$, then $J_1 \leq J_0$ according to Theorem \ref{the1}.
	\item If $\sum_{i,j} G_t^1(i,j) = \sum_{i,j} G_t^0(i,j)$, KM algorithm ensures to find the maximum weight of $O_t$, which means $J_1$ is the minimum, so we can derive that $J_1 \leq J_0$.
\end{enumerate}

So far, we have proved that $J_1 \leq J_0$. And when $G_t^0$ remains constant, the problem on the upper level can be solved by the MILP algorithm. The updated hydrogen energy dispatch is defined as $H_t^1$ and $J_2 = J(H_t^1, G_t^0)$. On the basis of the optimality preserving property of MILP, we can conclude that  $J_2 \leq J_1$, which means,
\begin{equation}
J_2 \leq J_1 \leq J_0
\end{equation}

We prove that the objective function $J$ is monotonically decreasing in one iteration of Algorithm \ref{alg:algorithm1} and $J$ is also bounded. Therefore, for a monotone bounded performance sequence of the MILP problem, it must converge to the optimum eventually. The proof is complete.
\endproof

\ifCLASSOPTIONcaptionsoff
  \newpage
\fi



%\bibliographystyle{IEEEtran}
\addtolength{\textheight}{-12cm}   % This command serves to balance the column lengths

%%%%%%%%%%%%%%%%%%%%%%%%%%%%%%%%%%%%%%%%%%%%%%%%%%%%%%%%%%%%%%%%%%%%%%%%%%%%%%%%

%%%%%%%%%%%%%%%%%%%%%%%%%%%%%%%%%%%%%%%%%%%%%%%%%%%%%%%%%%%%%%%%%%%%%%%%%%%%%%%%

%%%%%%%%%%%%%%%%%%%%%%%%%%%%%%%%%%%%%%%%%%%%%%%%%%%%%%%%%%%%%%%%%%%%%%%%%%%%%%%%

%%%%%%%%%%%%%%%%%%%%%%%%%%%%%%%%%%%%%%%%%%%%%%%%%%%%%%%%%%%%%%%%%%%%%%%%%%%%%%%%

\bibliographystyle{IEEEtran}
\bibliography{references}

\begin{thebibliography}{10}
\providecommand{\url}[1]{#1}
\csname url@samestyle\endcsname
\providecommand{\newblock}{\relax}
\providecommand{\bibinfo}[2]{#2}
\providecommand{\BIBentrySTDinterwordspacing}{\spaceskip=0pt\relax}
\providecommand{\BIBentryALTinterwordstretchfactor}{4}
\providecommand{\BIBentryALTinterwordspacing}{\spaceskip=\fontdimen2\font plus
\BIBentryALTinterwordstretchfactor\fontdimen3\font minus
  \fontdimen4\font\relax}
\providecommand{\BIBforeignlanguage}[2]{{%
\expandafter\ifx\csname l@#1\endcsname\relax
\typeout{** WARNING: IEEEtran.bst: No hyphenation pattern has been}%
\typeout{** loaded for the language `#1'. Using the pattern for}%
\typeout{** the default language instead.}%
\else
\language=\csname l@#1\endcsname
\fi
#2}}
\providecommand{\BIBdecl}{\relax}
\BIBdecl

\bibitem{dominkovic2016zero}
D.~F. Dominkovi{\'c}, I.~Ba{\v{c}}ekovi{\'c}, B.~{\'C}osi{\'c},
  G.~Kraja{\v{c}}i{\'c}, T.~Puk{\v{s}}ec, N.~Dui{\'c}, and N.~Markovska, ``Zero
  carbon energy system of south east europe in 2050,'' \emph{Applied energy},
  vol. 184, pp. 1517--1528, 2016.

\bibitem{jia2020review}
Q.-S. Jia and T.~Long, ``A review on charging behavior of electric vehicles:
  data, model, and control,'' \emph{Control Theory and Technology}, vol.~18,
  no.~3, pp. 217--230, 2020.

\bibitem{wang2019tcharge}
G.~Wang, F.~Zhang, and D.~Zhang, ``tcharge-a fleet-oriented real-time charging
  scheduling system for electric taxi fleets,'' in \emph{Proceedings of the
  17th Conference on Embedded Networked Sensor Systems}, 2019, pp. 440--441.

\bibitem{zhang2017second}
H.~Zhang, S.~J. Moura, Z.~Hu, W.~Qi, and Y.~Song, ``A second-order cone
  programming model for planning pev fast-charging stations,'' \emph{IEEE
  Transactions on Power Systems}, vol.~33, no.~3, pp. 2763--2777, 2017.

\bibitem{vazifeh2018addressing}
M.~M. Vazifeh, P.~Santi, G.~Resta, S.~H. Strogatz, and C.~Ratti, ``Addressing
  the minimum fleet problem in on-demand urban mobility,'' \emph{Nature}, vol.
  557, no. 7706, p. 534, 2018.

\bibitem{fang2016health}
H.~Fang, Y.~Wang, and J.~Chen, ``Health-aware and user-involved battery
  charging management for electric vehicles: Linear quadratic strategies,''
  \emph{IEEE Transactions on Control Systems Technology}, vol.~25, no.~3, pp.
  911--923, 2016.

\bibitem{moghaddass2019smart}
R.~Moghaddass, O.~A. Mohammed, E.~Skordilis, and S.~Asfour, ``Smart control of
  fleets of electric vehicles in smart and connected communities,'' \emph{IEEE
  Transactions on Smart Grid}, vol.~10, no.~6, pp. 6883--6897, 2019.

\bibitem{liu2017decentralized}
M.~Liu, P.~K. Phanivong, Y.~Shi, and D.~S. Callaway, ``Decentralized charging
  control of electric vehicles in residential distribution networks,''
  \emph{IEEE Transactions on Control Systems Technology}, vol.~27, no.~1, pp.
  266--281, 2017.

\bibitem{pflaum2017probabilistic}
P.~Pflaum, M.~Alamir, and M.~Y. Lamoudi, ``Probabilistic energy management
  strategy for ev charging stations using randomized algorithms,'' \emph{IEEE
  Transactions on Control Systems Technology}, vol.~26, no.~3, pp. 1099--1106,
  2017.

\bibitem{morstyn2020conic}
T.~Morstyn, C.~Crozier, M.~Deakin, and M.~D. McCulloch, ``Conic optimisation
  for electric vehicle station smart charging with battery voltage
  constraints,'' \emph{IEEE Transactions on Transportation Electrification},
  2020.

\bibitem{long2017multi}
T.~Long, J.-X. Tang, and Q.-S. Jia, ``Multi-scale event-based optimization for
  matching uncertain wind supply with ev charging demand,'' in \emph{2017 13th
  IEEE Conference on Automation Science and Engineering (CASE)}.\hskip 1em plus
  0.5em minus 0.4em\relax IEEE, 2017, pp. 847--852.

\bibitem{li2017data}
C.~Li, C.~Liu, K.~Deng, X.~Yu, and T.~Huang, ``Data-driven charging strategy of
  pevs under transformer aging risk,'' \emph{IEEE Transactions on Control
  Systems Technology}, vol.~26, no.~4, pp. 1386--1399, 2017.

\bibitem{liu2017dynamic}
S.~Liu and A.~H. Etemadi, ``A dynamic stochastic optimization for recharging
  plug-in electric vehicles,'' \emph{IEEE Transactions on Smart Grid}, vol.~9,
  no.~5, pp. 4154--4161, 2017.

\bibitem{ghosh2017control}
A.~Ghosh and V.~Aggarwal, ``Control of charging of electric vehicles through
  menu-based pricing,'' \emph{IEEE Transactions on Smart Grid}, vol.~9, no.~6,
  pp. 5918--5929, 2017.

\bibitem{koufakis2019offline}
A.-M. Koufakis, E.~S. Rigas, N.~Bassiliades, and S.~D. Ramchurn, ``Offline and
  online electric vehicle charging scheduling with v2v energy transfer,''
  \emph{IEEE Transactions on Intelligent Transportation Systems}, vol.~21,
  no.~5, pp. 2128--2138, 2019.

\bibitem{zheng2018novel}
Y.~Zheng, Y.~Shang, Z.~Shao, and L.~Jian, ``A novel real-time scheduling
  strategy with near-linear complexity for integrating large-scale electric
  vehicles into smart grid,'' \emph{Applied Energy}, vol. 217, pp. 1--13, 2018.

\bibitem{tucker2019online}
N.~Tucker and M.~Alizadeh, ``An online admission control mechanism for electric
  vehicles at public parking infrastructures,'' \emph{IEEE Transactions on
  Smart Grid}, vol.~11, no.~1, pp. 161--170, 2019.

\bibitem{zhang2020power}
H.~Zhang, Z.~Hu, and Y.~Song, ``Power and transport nexus: Routing electric
  vehicles to promote renewable power integration,'' \emph{IEEE Transactions on
  Smart Grid}, 2020.

\bibitem{uber}
``Uber[online],'' \url{https://www.uber.com.cn/}.

\bibitem{didi}
``Didi[online],'' \url{https://www.didiglobal.com/}.

\bibitem{zhang2019joint}
H.~Zhang, C.~J. Sheppard, T.~E. Lipman, and S.~J. Moura, ``Joint fleet sizing
  and charging system planning for autonomous electric vehicles,'' \emph{IEEE
  Transactions on Intelligent Transportation Systems}, 2019.

\bibitem{sarkar2011mw}
S.~Sarkar and V.~Ajjarapu, ``Mw resource assessment model for a hybrid energy
  conversion system with wind and solar resources,'' \emph{IEEE transactions on
  sustainable energy}, vol.~2, no.~4, pp. 383--391, 2011.

\bibitem{ulleberg2003modeling}
{\O}.~Ulleberg, ``Modeling of advanced alkaline electrolyzers: a system
  simulation approach,'' \emph{International journal of hydrogen energy},
  vol.~28, no.~1, pp. 21--33, 2003.

\bibitem{zhang2016optimal}
H.~Zhang, Z.~Hu, Z.~Xu, and Y.~Song, ``Optimal planning of pev charging station
  with single output multiple cables charging spots,'' \emph{IEEE Transactions
  on Smart Grid}, vol.~8, no.~5, pp. 2119--2128, 2016.

\bibitem{bochet2012balancing}
O.~Bochet, R.~Ilk{\i}l{\i}{\c{c}}, H.~Moulin, and J.~Sethuraman, ``Balancing
  supply and demand under bilateral constraints,'' \emph{Theoretical
  Economics}, vol.~7, no.~3, pp. 395--423, 2012.

\bibitem{west1996}
D.~B. West, ``Introduction to graph theory,'' \emph{Upper Saddle River, NJ:
  Prentice hall}, vol.~2, 1996.

\bibitem{SANY}
``Sany se13122.[online],'' \url{http://www.sanyhi.com/}.

\bibitem{wind}
``The national meteorological information center.[online],''
  \url{http://data.cma.cn/en}.

\bibitem{liuphd2017}
L.~Pengfei, ``Study on capacity optimal configuration and energy management of
  integrated wind-solar-hydrogen-storage power supply system,'' Ph.D.
  dissertation, Zhejaing University, 2017.

\bibitem{Zhuphd2017}
Z.~Jimin, ``Performance simulation and energy management of hybrid wind-pemfc
  power generation system,'' Ph.D. dissertation, Shandong University, 2017.

\bibitem{zhang2015integrated}
H.~Zhang, Z.~Hu, Z.~Xu, and Y.~Song, ``An integrated planning framework for
  different types of pev charging facilities in urban area,'' \emph{IEEE
  Transactions on Smart Grid}, vol.~7, no.~5, pp. 2273--2284, 2015.

\bibitem{rahman2016review}
I.~Rahman, P.~M. Vasant, B.~S.~M. Singh, M.~Abdullah-Al-Wadud, and N.~Adnan,
  ``Review of recent trends in optimization techniques for plug-in hybrid, and
  electric vehicle charging infrastructures,'' \emph{Renewable and Sustainable
  Energy Reviews}, vol.~58, pp. 1039--1047, 2016.

\bibitem{evdata}
``Taxis - smart city research group[online],''
  \url{https://www.cse.ust.hk/scrg/}.

\bibitem{stats}
``National bureau of statistics[online],'' \url{http://www.stats.gov.cn/}.

\bibitem{johnson1962connectivity}
D.~M. Johnson, A.~L. Dulmage, and N.~S. Mendelsohn, ``Connectivity and
  reducibility of graphs,'' \emph{Canadian Journal of Mathematics}, vol.~14,
  pp. 529--539, 1962.

\end{thebibliography}
\begin{IEEEbiography}[{\includegraphics[width=1in,height=1.25in,clip,keepaspectratio]{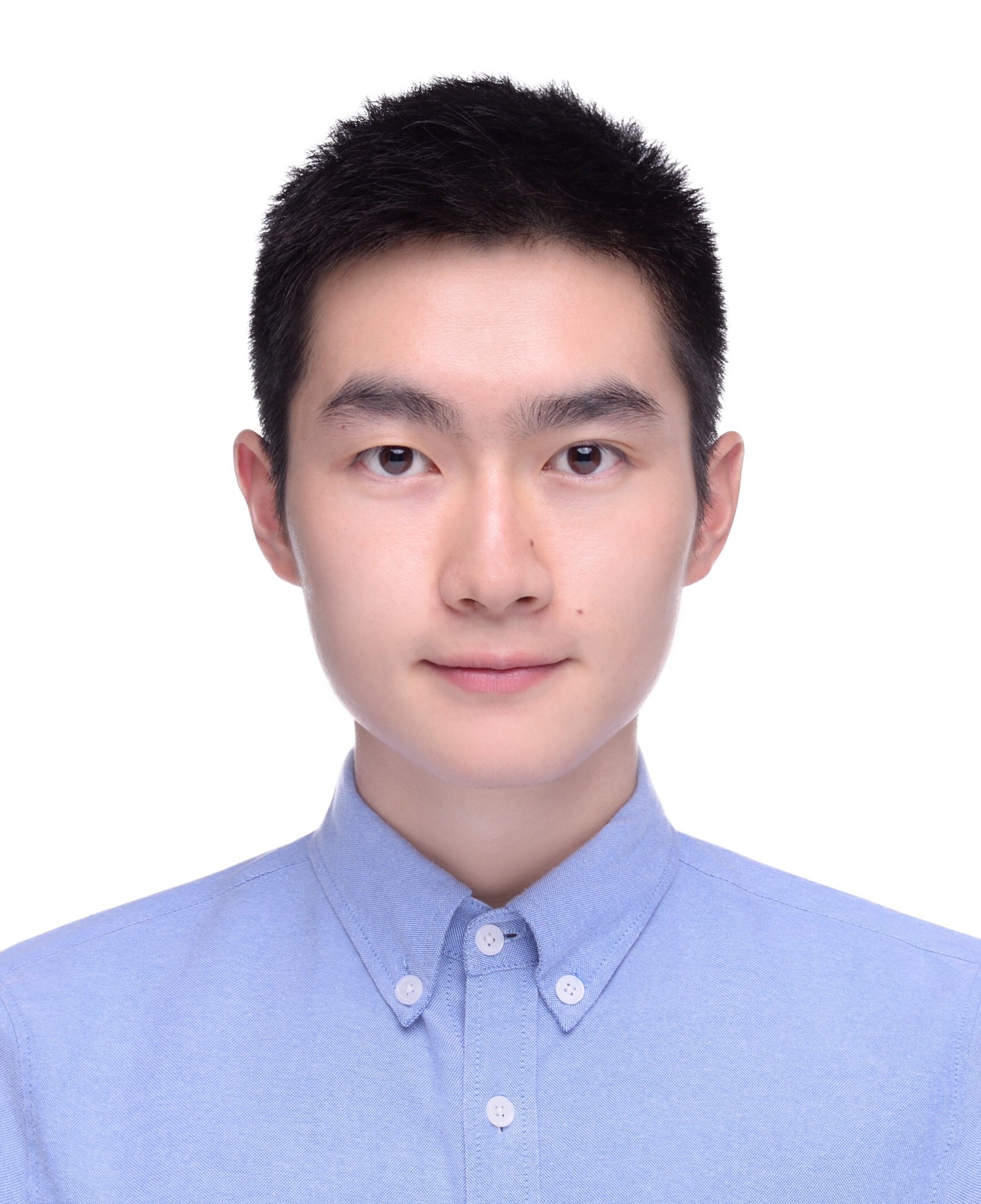}}]{Teng Long}
	(S'17) received the B.S. degree in automation from Tsinghua University, Beijing,
	China, in 2017. He is currently pursuing the Ph.D. degree with the Center for Intelligent and Networked Systems, Department of Automation, Tsinghua University, Beijing, China. 
	
	His current research interests include energy management of smart grid, event-based optimization, and large-scale optimization problem.
\end{IEEEbiography}

\begin{IEEEbiography}[{\includegraphics[width=1in,height=1.25in,clip,keepaspectratio]{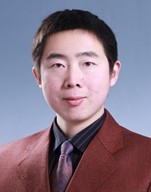}}]{Qing-Shan Jia}
	(S'02-M'06-SM'11) received the B.S. degree in automation and the Ph.D. degree in control science and engineering from Tsinghua University, Beijing, China, in 2002 and 2006, respectively. He was a Visiting Scholar at Harvard University, in 2006, Hong Kong University of Science and Technology, in 2010, and at Massachusetts Institute of Technology, in 2013. He is currently an Associate Professor at the Center for Intelligent and Networked Systems, Department of Automation, BNRist, Tsinghua University.
	
	His current research interests include theories and applications of discrete event dynamic systems and simulation-based performance evaluation and optimization of complex systems.
\end{IEEEbiography}
\end{document}